\documentclass[reprint,amsmath,amssymb,amsfonts,aps,prb,superscriptaddress]{revtex4-2}
\usepackage{graphicx}
\usepackage{bm}
\usepackage{epsfig}
\usepackage{epstopdf}
\usepackage{xcolor}
\usepackage{hyperref}
\hypersetup{colorlinks=true,allcolors=blue}
\usepackage{natbib}
\usepackage{mathtools}
\usepackage{siunitx}
\usepackage{soul,xcolor}
\usepackage{amsmath,amssymb}

\setstcolor{red}
\DeclareUnicodeCharacter{2212}{-}
\DeclarePairedDelimiterXPP\BigOSI[2]%
  {\mathcal{O}}{(}{)}{}%
  {\SI{#1}{#2}}

\begin{document}
%\date{\today}
\title{    Strong Disorder Renormalization  Group Method for 
Bond Disordered Antiferromagnetic Quantum Spin Chains with Long Range Interactions: Excited States and Finite Temperature Properties
}

\author{S. Kettemann}
\email[]{s.kettemann@jacobs-university.de}

\affiliation{
Max Planck Institute for the Physics of Complex Systems, 01187 Dresden, Germany}

\affiliation{Department of Physics and Earth Sciences and 
Department of Computer Science, Constructor University, Bremen 28759, Germany}

\begin{abstract} 
We extend the recently introduced  strong disorder renormalization group method in real space,
well suited to study bond disordered antiferromagnetic power law coupled quantum spin chains,
to study excited states, and 
finite temperature  properties. 
 First, we  apply it to 
 a short range  coupled spin chain,  which is defined by  the  model with  power law interaction,   keeping only interactions  between   adjacent spins. 
  We show that the distribution of the absolute value of the 
   couplings is  the infinite randomness fixed point distribution. However, the sign of the couplings becomes 
    distributed, and the number of negative couplings  increases with temperature $T.$
  
Next,  we derive the 
  Master equation 
  for the power law  long  range interaction between all spins with power exponent $\alpha$. While  the sign of the couplings is found to be 
    distributed, the distribution of the 
     coupling amplitude is given by the strong disorder
      distribution with finite width $2\alpha,$
      with small corrections for $\alpha >2$. 
      
  Resulting  finite temperature properties 
of both short and power law long ranged spin systems are derived,
 including the magnetic susceptibility, concurrence and 
  entanglement entropy. 
\end{abstract}

\maketitle

\section{Introduction}
The  magnetic  properties of a wide range of    materials
 are dominated by randomly placed 
 quantum spins 
    which are coupled by  long range interactions
\cite{anderson58,loehneysen,Kettemann2023,Kettemann2024,reviewtls,Yu}.
   Meanwhile, one can realize 
 tunable interactions   between atoms trapped near photonic crystals 
 \cite{grass} and by coupling  Rydberg states with opposite parity \cite{Signoles2021,Brow2020}. 
 Trapped ions with power-law interactions have  been realized\cite{Islam2013,Richerme2014}.
Moreover,  single nitrogen-vacancy centers in diamond allow to probe the dynamics of 
     disordered spin ensembles at a diamond surface
               \cite{Lukin2022,Davis2022}.

  It remains, however,  a challenge to derive  the 
  thermodynamic and dynamic  properties of such systems. 
 The  strong disorder renormalization group  (SDRG) method
  has been introduced to study
disordered   quantum spin chain models\cite{bhattlee81,bhattlee82,fisher94,fisher95,monthus,Igloi2018}.
 Short range antiferromagnetically coupled disordered spin chain models 
are found to be governed by the  infinite randomness fixed point (IRFP). 
Their  ground state   is 
    composed of  a  product state of randomly placed 
    pairs of spins in 
     singlet states, the {\it random singlet phase}.  The SDRG method has also been applied to 
 other   short range random  quantum  spin chains, like the transverse field  Ising model\cite{monthus}.
      %It allows one  to derive their  thermodynamic properties\cite{monthus} and  dynamical properties like   entanglement dynamics\cite{Vosk2013,Vosk2014,igloi2012}. 
         Recently, 
          the SDRG  method
          was extended
          to  study disordered  spin   $S=1/2-$chains
         with antiferromagnetic  power law  long range XX-interactions.
          The 
           ground state  was shown 
           for sufficiently large power
           exponent $\alpha$
           to be  still   composed    of  
           a product state of random singlets, but 
           with a fixed point distribution of finite width\cite{Moure2015,Moure2018,Mohdeb2020,Kettemann2025}.
           This was confirmed by 
          numerical exact diagonalization  and  extensions of the 
            DMRG method\cite{Mohdeb2020}.  A similar  fixed point distribution  was  found for long range coupled 
           transverse field Ising chains
           \cite{Juhasz2014,Juhasz2016}.
          
      It remains to derive 
excited states and finite temperature 
properties of long range bond disordered antiferromagnetic 
quantum anisotropic Heisenberg spin chains, which correspond to 1-dimensional systems of   interacting  Fermions with disordered  long range hopping and interactions\cite{Moure2015}.
 To this end, we introduce here a real space representation of   the  SDRG-X method 
to study properties of excited  states 
\cite{Pekker2014,Huang2014,Aram2024}, which 
 was recently applied to study excitations 
            \cite{Mohdeb2022}  
and entanglement  dynamics  after global quantum quenches of 
 long range coupled spin chains
\cite{Mohdeb2023}.             

\begin{figure}
    \includegraphics[scale=0.35]{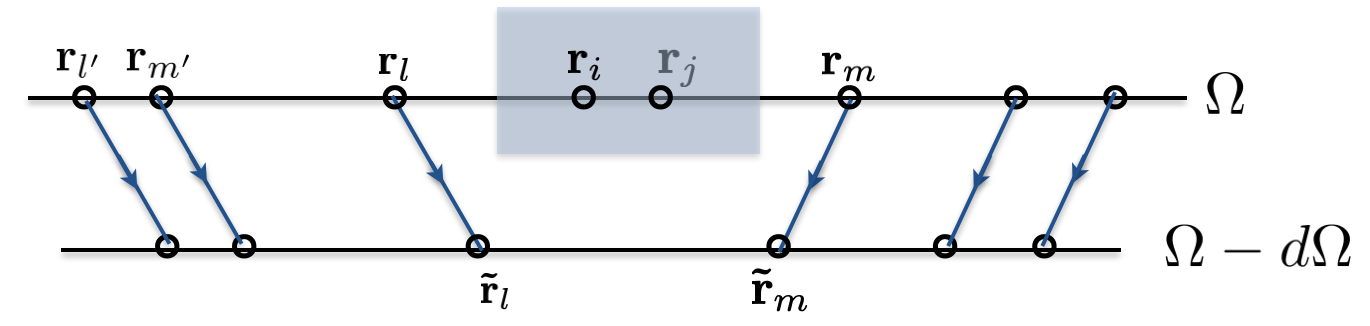}
\vspace*{-.5cm}\caption{Strong 
disorder RG step
for bond disordered short range coupled spin chains: Decimation of strongest coupled spin pair $(i,j)$,
highlighted by the shaded area, 
whose coupling defines the  RG scale $\Omega.$ It
is followed by renormalization of the positions of spins,  ${\bf r}_{l} \rightarrow \tilde{{\bf r}}_{l}$ and a reduction of the RG scale to $\Omega - d\Omega.$ } 
    \label{RGnn}
\end{figure}

\begin{it}
Model.\end{it}— 
We study long range antiferromagnetic XX-coupled
quantum spin chains  of $N$
$S=1/2-$spins, 
randomly placed at positions ${\bf r}_i$ 
on a line
of length $L,$ 
as shown  in Fig. \ref{RGnn} 
and defined by 
 \begin{equation}\label{H}
H=\sum_{i<j} J_{ ij}\left(S_{i}^{x}\,S_{j}^{x}+S_{i}^{y}\,S_{j}^{y}\right).
\end{equation}
We take   $N$ to be even and take open boundary conditions. 
  The couplings between spins at sites $i,j$ are antiferromagnetic and long-ranged, decaying with a power law  with distance $r_{ij} = |{\bf r}_i-{\bf r}_j|$ with  exponent $\alpha$, 
  \begin{equation} \label{jcutoff}
J_{ij} =
J_0\left|({\bf r}_i-{\bf r}_j)/a\right |^{-\alpha},
\end{equation}
with  $J_0 >0,$ and the condition that  $|{\bf r}_i-{\bf r}_j| > a$
for all $i,j,$ where $a \ll L/N$ is the smallest possible distance between spins. 

    It is insightful to use     the Jordan-Wigner transformation which maps    the spin chain  Eq. (\ref{H}) onto
    the Hamiltonian of  fermions \cite{Moure2015,Kettemann2025}.
  One finds that 
   the long range interactions introduce dynamic phase   correlations,
    which make the problem a many body problem, even for the XX-coupled spin chains.

\section{SDRG-X Method}

\begin{figure}
    \includegraphics[scale=0.18]{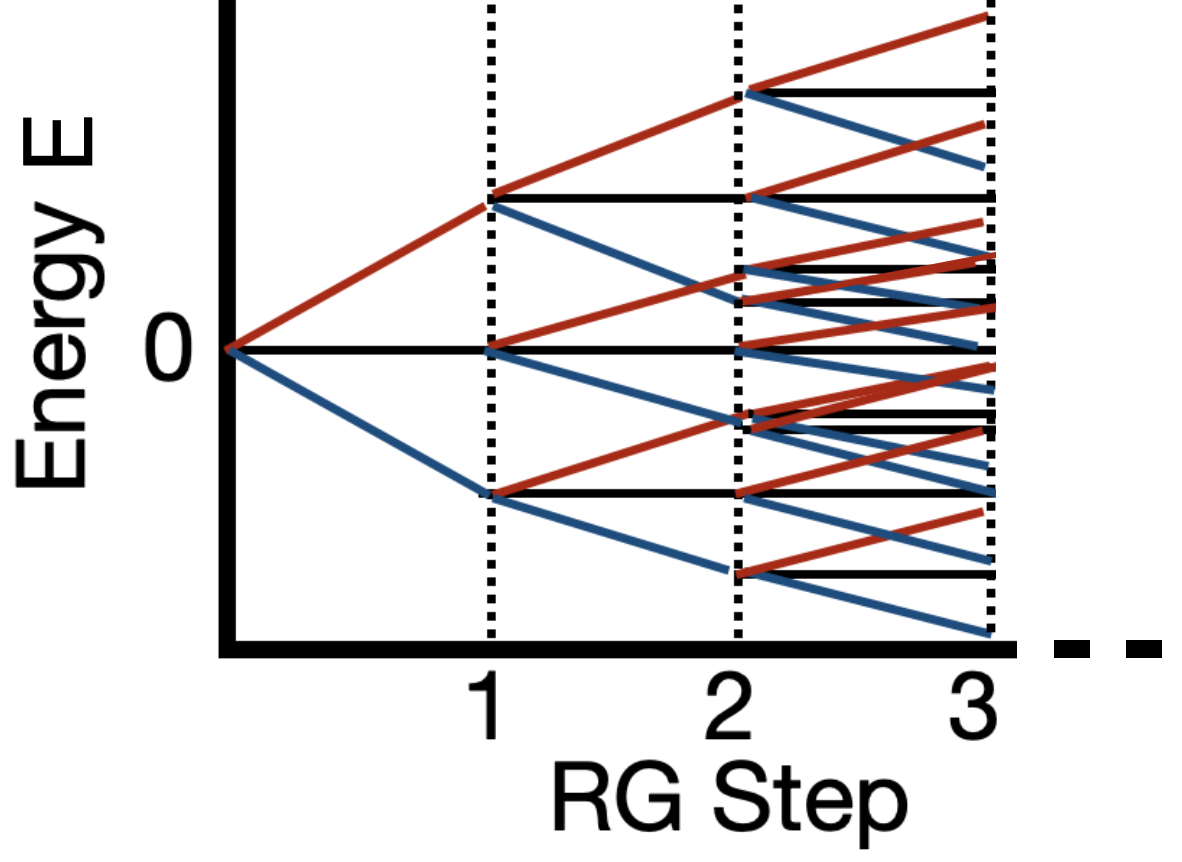}
\vspace*{-.0cm}\caption{Schematic SDRG-X procedure:
 at each RG step four possible pair states
  are indicated by blue, black and red lines ( with pair energies $E=-J/2,0,0,+J/2$, respectively). 
  After
  $N/2$ RG steps the  many body eigenstates with total energy E is obtained, following a specific SDRG path.  } 
    \label{RGscheme}
\end{figure}

We review the SDRG-X method,
 as introduced in Ref. \cite{Pekker2014}.
Assuming that all many body Eigen states of the Hamiltonian Eq. (\ref{H}) can be written 
 in good approximation as tensor products of pair states, one  starts   by identifying the strongest coupled pairs of spins $(i,j)$ for a given initial distribution of couplings $P(J,\Omega_0),$
 where $\Omega_0$ is that largest energy scale. 
Rewriting the Hamiltonian as $\hat{H} = \hat{H}_0 + \hat{V},$
where  the Hamiltionian of the most strongly coupled 
 pair of spins  $(i,j)$ is given by $\hat{H}_0  =J^x_{ ij}\left(S_{i}^{x}\,S_{j}^{x}+S_{i}^{y}\,S_{j}^{y}\right)$ and  $\hat{V}$ models the interaction 
 of spins $(i,j)$
 with   all other $N-2$ spins and between them. 
   Diagonalising  $\hat{H}_0,$ one finds
   its  four Eigenstaes $|s \rangle,$
   with $s \in \{0,1,2,3\}$ with Eigen energies
   $E_{s}.$
Projecting next that pair on one of the pair states $s,$
 we construct the effective Hamiltionan of the remaining $N-2$ spins 
 $\hat{H}_{\rm eff},$  such that 
 it commutes with $\hat{H}_0$ and is therefore diagonal in its Eigenstates $|s \rangle.$
  Then, the effective Hamiltonian can be written as
\begin{equation}
 \hat{H}_{\rm eff} = \exp( \i \hat{S}) \hat{H}
 \exp(- \i \hat{S}).
\end{equation}
We expand $\hat{H}_{\rm eff}$  to 2nd order in  $\hat{S}$ and enforce 
 commutation  with  $\hat{H}_0$.
  Denoting subspaces $D_u$
   such that all states $s\in D_u$
have the same Eigene energy $E_s,$
  we 
  find that  $\hat{H}_{\rm eff}$  is to 2nd order 
   in  $\hat{V}$ given by 
  \begin{eqnarray} \label{heff}
 \hat{H}_{\rm eff} & \approx & 
 \sum_u
 \sum_{s,s' \in D_u} |s \rangle \langle s' |
 [ E_s \delta_{s,s'} +
\langle s | \hat{V} |s' \rangle 
+ \frac{1}{2} \sum_{s'' \notin D_u}  
\nonumber \\
 &&\langle s | \hat{V} |s'' \rangle \langle s'' | \hat{V} |s' \rangle (\frac{1}{E_{s}-E_{s''}}
 +\frac{1}{E_{s'}-E_{s''}})
 ].
\end{eqnarray}
The last terms define  the  renormalized couplings and local fields for all 
 remaining $N-2$ spins. 
 Even  if the form of the Hamiltionian remains 
   unchanged, 
 the  renormalized couplings may  differ from the initial ones,  so that the distribution function of couplings is changed to 
 $P_E(J,\Omega_0 - d\Omega),$
  where $\Omega_0 - d\Omega$ is the 
 largest energy scale in the reduced system of $N-2$
 spins. $E$ is the total energy of the system.
 Repeating this procedure until all $N$
 spins formed pairs we find 
  the distribution of  effective couplings $P_E(J,\Omega)$
  in the limit of $\Omega \rightarrow 0$, which allows to derive thermodynamic and dynamic properties of the spin chain.
Its  total energy $E$
  is obtained by summing over all pair energies $E_{s_n},$ as obtained in the $n-$ th RG step, 
  $n\in \{1,2,.,N/2 \},$ yielding
  $E= \sum_{n=1}^{N/2} E_{s_n}.$
  Thus, a specific  total energy $E$
  corresponds to a specific RG path, as sketched in Fig. \ref{RGscheme}.

  Instead of using a microcanonical ensemble at energy $E,$
  it is often  more convenient to consider a canonical ensemble 
  at bath temperature $T.$
Then, any of the four  
     pair states $s \in \{ 0,...,3\}$  
     of the strongest coupled pair  can be occupied
     at RG scale  $\Omega$     
     with probability
     \begin{equation} \label{occupation}
     p_{s} (E_{s} (\Omega),T) = \exp(- \beta E_{s}(\Omega \sigma))/Z(\Omega),
     \end{equation}
     where
     $\beta = 1/(k_{\rm B} T)$ and
      the partition sum is given by 
      $Z(\Omega)= \sum_{s} \exp(- \beta E_{s} (\Omega)).$

\section{
Nearest Neighbour  Coupling}

We  first apply   the  SDRG-X
             procedure 
             to
             study the excited states 
              and finite temperature properties of 
        randomly placed spins on a chain, 
   with power law XX-coupling,   Eq. (\ref{H}),   
          keeping only the coupling between adjacent spins. 
            We identify  the strongest coupling  
              $J_{ij} = \Omega,$
              as  highlighted
               in the  example of randomly placed spins
 in Fig.  \ref{RGnn}.
While   the ground state of the 
               spin pair $(i,j)$ is  the singlet state $|0_{ij}\rangle,$ 
               the excited states are one of 
                three triplet states, 
                 the two unentangled states $|1_{ij} \rangle =|\uparrow_i\,\uparrow_i\rangle$ and $|2_{ij} \rangle =|\downarrow_i\,\downarrow_j\rangle$,
                 and the entangled triplet state
$|3_{ij} \rangle =|\left(|\uparrow\,\downarrow\rangle+|\downarrow\,\uparrow\rangle\right)/\sqrt{2}.$   
   The corresponding 
   Eigenergies of  spin pair $i,j$
   are  given by $E_0=-J_{ij}/2,$
   $E_1=E_2=0,$ and $E_3=J_{ij}/2.$

   \subsection{SDRG-X Rules}
  
  Insertion into Eq. (\ref{heff}) yields 
  no local fields,
  $\langle s | \hat{V} |s \rangle = 0$,
   for all states $s \in \{0,1,2,3\}.$
But a new coupling is generated  between those spins 
    adjacent to $i,j$,  denoted by $l,m$ in Fig.  \ref{RGnn}.
 Thereby, we find   the following renormalization  rules:

{\it Singlet State.—}
If $ (i,j) $ is in the 
singlet state, $s=0,$ $\left(|\uparrow\,\downarrow\rangle-|\downarrow\,\uparrow\rangle\right)/\sqrt{2},$  then  the generated  coupling is given by 
\begin{equation}
\label{RGe}
     \tilde{J}_{lm}=  \frac{J_{i l} J_{jm}}{J_{ij}}.
\end{equation}
            
{\it Unentangled Triplet States.—}
              If  the pair $ (i,j)   $ is projected onto one of the unentangled
               triplet states, $s=1,$ $|\uparrow\,\uparrow\rangle$ or 
               $s=2,$  $|\downarrow\,\downarrow\rangle,$
 a coupling between spins at sites $l,m$ is generated 
 with 
\begin{equation}
\label{RGue}
    \tilde{J}_{lm}= - \frac{J_{i l} J_{jm}}{J_{ij}}.
\end{equation}
Thus,  the coupling acquires a minus sign, generating ferromagnetic couplings, even though 
 all original couplings are antiferromagnetic.

{\it Entangled Triplet  State.—}
If $ (i,j) $ is in the entangled tripled state, $s=3,$ $\left(|\uparrow\,\downarrow\rangle+|\downarrow\,\uparrow\rangle\right)/\sqrt{2}$ or in the singlet state, then  the generated  coupling is 
\begin{equation}
\label{RGe}
     \tilde{J}_{lm}=  \frac{J_{i l} J_{jm}}{J_{ij}}.
\end{equation}
Thus, while the amplitude of the generated couplings is independent on the  state which the paired spins 
 form,  the nonentangled states generate a different sign of the coupling. Thus, we can summarize all 
 SDRG rules as
 \begin{equation}
\label{RGall}
     \tilde{J}_{lm}= \sigma_{s} \frac{J_{i l} J_{jm}}{J_{ij}},
\end{equation}
where $\sigma_{s}=1$ for $s=0,3$
and $\sigma_{s}=-1$ for $s=1,2.$
 
Since  the magnitude of the adjacent  couplings depends on their distances,  
it is convenient to  represent  the renormalized  couplings $\tilde{J}$
 in terms of renormalized distances $\tilde{r}$ as  
 $\tilde{J}_{lm}=J_0/\tilde{r}_{lm}^{\alpha}.$
Now, the RG rules can be recast for the  renormalized distance $\tilde{r}_{lm} = \tilde{r},$
as sketched  in Fig. \ref{RGnn}.
At the RG length scale $\rho = (\Omega/J_0)^{-1/\alpha}$
the renormalized distance is thus 
     \begin{eqnarray} \label{reff}
                &&\tilde{r} =
               \frac{r_{li} r_{jm}}{\rho}.
              \end{eqnarray}     
              In order to trace the sign of the coupling 
              $\sigma,$
              we introduce the
 distribution function of distances  $P_{\sigma}(\tilde{r},\Omega)$ with sign of coupling $\sigma.$ 
 This  
 is related to
 the distribution of couplings with amplitudes 
 $\tilde{J} \le \Omega$  having  sign $\sigma,$
 $P_{\sigma}(\tilde{J},\Omega)$  by 
 $P_{\sigma}(\tilde{r},\Omega) = (\alpha \tilde{J}/\tilde{r}) P_{\sigma}(\tilde{J},\Omega)|_{\tilde{J}=\Omega_0 \tilde{r}^{-\alpha} }.$  
  The normalization condition
  is  given by
  $\sum_{\sigma = \pm} \int_0^{\Omega} d \tilde{J}   P_{\sigma}(\tilde{J},\Omega)=1.$
  Accordingly, the normalization for the 
   distribution of distances is given by 
   $\sum_{\sigma = \pm} \int_{\rho}^{\infty} \tilde{r}  d P_{\sigma}(\tilde{r},\Omega)=1.$

    \subsection{Finite Temperature
    Distribution Function} 
 The  distribution functions  for excited states  can be  derived from a generalized Master equation,
  employing the real space formulation introduced in Ref.\cite{Kettemann2025}.
 To obtain that Master equation, we note that 
when the  
spin pair  $(i,j)$  forms a state $|s \rangle$
 at RG scale $\Omega = J_{ij},$
 the  couplings between the spin pairs
  $(l,i)$ and $(j,m)$ are taken away
   while a coupling between 
    spin pair $(l,m)$ is newly created with 
    renormalized coupling $\tilde{J}_{lm}.$
     In the representation of distances 
      this corresponds to take 
      the edges between spin pairs
  $(l,i)$ and $(j,m)$ 
  with distances  $r_{l,i}=R_L$ and $r_{j,m}=R_R$
  away and to create an edge between spin pair 
  $(l,m)$ with renormalized distance $\tilde{r}_{lm}=\tilde{r},$
  as shown    in Fig.  \ref{RGnn}, thereby 
replacing the bare distance 
${r}_{lm}={r}.$

We derive 
the   Master equation 
for  the distribution function of distances at temperature $T$, 
 $P_{\sigma,T}(\tilde{r},\Omega)$ 
for the short ranged model  in Appendix A, where we  find
\begin{eqnarray} \label{mnn}
 &&-\frac{d}{d\Omega}  P_{\sigma,T}(\tilde{r},\Omega) =
  \sum_{\sigma' } P_{\sigma',T}(\Omega,\Omega) 
  \sum_{\sigma_L = \pm}
  \int_{\rho}^{\infty} dR_L  \sum_{\sigma_R = \pm} 
 \nonumber \\
 &&  \int_{\rho}^{\infty}
 dR_R 
 P_{\sigma_L,T}(R_L,\Omega) P_{\sigma_R,T}(R_R,\Omega) 
  \sum_{s=0}^3   
        \nonumber \\
 &&       p_{s} (E_{s} ( \sigma' \Omega),T) 
 \delta (\tilde{r}-  \frac{R_L R_R}{\rho}) \delta_{\sigma,\sigma_L \sigma_R \sigma' \sigma_{s} }.
      \end{eqnarray}  
      Here, the state with energy $E_{s}(\sigma \Omega)$  at RG scale $\Omega$ with coupling sign $\sigma$
     is  occupied with probability Eq. (\ref{occupation})
     \begin{equation} \label{occupation}
     p_{s} (E_{s} (\Omega),T) = \exp(- \beta E_{s}(\Omega \sigma))/Z(\Omega),
     \end{equation}
     where 
      the partition sum is given by 
      $Z(\Omega)=  2 + 2 \cosh (\Omega/2).$
We  make  the product Ansatz 
\begin{equation} \label{ansatz}
P_{\sigma,T}(\tilde{r},\Omega) = P_{\sigma,T}(\Omega)
P_T(\tilde{r},\Omega),
\end{equation}
 where $P_{\sigma,T} (\Omega)$ is the probability 
 at temperature $T$
 that the sign of any  coupling $\tilde{J}\le \Omega$
is $\sigma.$ 

Let us first  find the distribution $P_T(\tilde{r},\Omega) =\sum_{\sigma =\pm} P_{\sigma,T}(\tilde{r},\Omega),$  summing over the sign of couplings $\sigma$. 
Performing the integral over one of the distances $R_L$, and with the  normalization
  condition 
$ \sum_{\sigma =\pm} P_{\sigma,T} (\Omega) =1$ 
it follows
 that it  
 is governed for $d \Omega \rightarrow 0$
 by the Master equation 
 \begin{eqnarray} \label{mnn}
 &&-\frac{d}{d\Omega}  P_T(\tilde{r},\Omega) =
 P_T(\Omega,\Omega) 
\int_{\rho}^{\infty}
 dR_R  \frac{\rho}{R_R} \times
 \nonumber \\
&& P_T(R_R,\Omega) P_T(\rho \tilde{r}/R_R,\Omega).
      \end{eqnarray}  
This is the same  Master equation, as was previously derived for the 
    distribution of distances in the ground state
     of the short ranged AFM coupled spin chain\cite{Kettemann2025}.
 It is  independent of temperature $T,$
and we recover the infinite disorder fixed point distribution
\begin{equation} \label{ptrIRFP}
P_T(\tilde{r},\Omega) =  \frac{c(\Omega)}{\tilde{r}} (\frac{\rho}{\tilde{r}})^{c(\Omega)}   ~ \theta(\tilde{r}/\rho-1),
\end{equation}
 with $\theta(x)$  the unit step function,
   $\theta(x>0) =1$, and   $\theta(x<0) =0,$
 and  $c(\Omega) = \alpha/\Gamma_{\Omega},$
with $\Gamma_{\Omega} = \ln (\Omega_0/\Omega).$
   Transforming back to the distribution function of couplings
  $ P(\tilde{J},\Omega) = \tilde{r}/(\alpha \tilde{J})  P(\tilde{r},\Omega)|,$
  we find 
  \begin{equation} \label{psdrg}
  P_T(\tilde{J},\Omega) = 
  \frac{1}{\Omega \Gamma_{\Omega}}(\frac{\Omega}{\tilde{J}})^{1-1/\Gamma_{\Omega}} 
  \theta(\Omega/\tilde{J}-1).
  \end{equation}
  which is the infinite randomness
   fixed point distribution function 
 with width  $\Gamma_{\Omega} = \ln (\Omega_0/\Omega),$
 diverging to infinity for $\Omega \rightarrow 0.$
 Here, we find that it applies to
the  excited states
  of the XX-chains, as well, 
 and is independent of temperature $T.$
Thus, with the Ansatz for 
$ P_{\sigma,T}(\tilde{r},\Omega),$   given by  Eq. (\ref{ansatz}) 
it remains only to find  the Bernoulli distribution 
for the sign of couplings
$P_{\sigma,T} (\Omega).$ Insertion of  Eq. (\ref{ansatz}) 
with the solution for $P_T(\tilde{r},\Omega),$ 
given by Eq. (\ref{ptrIRFP}), yields the Master equation for 
$P_{+,T} (\Omega)$ as 
 \begin{eqnarray} \label{mpst}
&& \frac{d}{d \Omega} P_{+,T} (\Omega) 
\nonumber \\
&=& g_{\alpha} (\Omega) (1-
p_{+,T} (\Omega) P_{+,T} (\Omega)^2) ( 2P_{+,T} (\Omega)-1).
  \end{eqnarray}
   Here, we  defined the function 
   \begin{equation} \label{galpha}
g_{\alpha} (\Omega) = \frac{1}{\Omega} (\frac{1}{2 \alpha}- \frac{1}{\ln \Omega_0/\Omega}).
  \end{equation}
 The   distribution 
 $p_{+,T}$ is the 
 probability, that the sign is not switched during an RG step, which is equal to the
 sum of occupation probabilities Eq. 
  (\ref{occupation}) of the two states, which do not yield a sign change under renormalization, 
  \begin{equation}
  p_{+,T} (\Omega)= \cosh (\Omega/(2 T))/(1+\cosh (\Omega/(2 T))).
  \end{equation}
For large RG scales $\Omega$ we thus find  $p_{+,T} (\Omega \gg T)=1.$  
 In   the 
  high temperature limit, on the other hand, 
  all states are occupied with equal probability  $p_{+,T} (\Omega \ll T)=1/2$.
  We know that 
initially all couplings are antiferromagnetic, so that
  $P_{+,T} (\Omega \gg T) =1$.
  As the RG scale is lowered to  $\Omega < T,$
  the sign changes can occur more frequently, when triplet states become  occupied. 
When  $\Omega \rightarrow 0,$  both signs of the couplings  
 become   equally likely, 
$P_{+,T} (\Omega \rightarrow 0) =1/2.$
Indeed, we find that Eq. 
(\ref{mpst}) has 
for $\Omega \le 2T$ the solution for $\sigma \in \{+,-\} $
\begin{equation} \label{pst}
    P_{\sigma,T} (\Omega) = \frac{1}{2} + \sigma \frac{1}{2}
    \left( \frac{\Omega}{2T} \right)^{7/(8\alpha)}
     \left( \frac{\ln(\Omega_0/\Omega)}{\ln(\Omega_0/(2T)} \right)^{7/4},
\end{equation}
while $P_{+,T} (\Omega)  =1$ for $\Omega \ge 2T,$
 $P_{-,T} (\Omega)  =0$ for $\Omega \ge 2T,$

Thus, we find that the distribution of couplings 
smaller than RG scale $\Omega$
is given by 
  \begin{equation} \label{psdrgf}
  P_{\sigma,T}(\tilde{J},\Omega) =  P_{\sigma,T} (\Omega)
  \frac{1}{\Omega \Gamma_{\Omega}}(\frac{\Omega}{\tilde{J}})^{1-1/\Gamma_{\Omega}} 
  \theta(\Omega/\tilde{J}-1),
  \end{equation}
where $ P_{\sigma,T} (\Omega)$ as given by Eq. (\ref{pst}) is the probability that the coupling 
$\tilde{J} <\Omega$ has the sign $\sigma.$
 For $\Omega \rightarrow 0$  that probability approaches 
 $ P_{\sigma,T} (\Omega \rightarrow 0) \rightarrow 1/2.$

  Having the full  distribution 
  function of couplings 
  Eq. (\ref{psdrgf}),
  allows us  to derive  finite temperature  properties of this model.
 Since the spectrum of the pair states of XX coupled spins  is 
 symmetric in a sign change of exchange coupling, see Fig. \ref{RGscheme}, 
the sign of the coupling does not enter physical properties in that case,
so that  it is sufficient to have 
 the distribution function  of coupling amplitudes 
  \begin{eqnarray} \label{psdrgabs}
  P_{T}(\tilde{J},\Omega) &=& \sum_{\sigma =\pm} 
  P_{\sigma,T}(\tilde{J},\Omega) 
  \nonumber \\
  &=&  
  \frac{1}{\Omega \Gamma_{\Omega}}(\frac{\Omega}{\tilde{J}})^{1-1/\Gamma_{\Omega}} 
  \theta(\Omega/\tilde{J}-1),
  \end{eqnarray}
  with $\Gamma(\Omega) = \ln \Omega_0/\Omega,$
  which is the infinite randomness fixed point distribution.  We have shown 
  that it is
   valid for this short ranged model for  any temperature $T.$

\section{
Long Range Coupling}

\begin{figure}
    \includegraphics[scale=0.18]{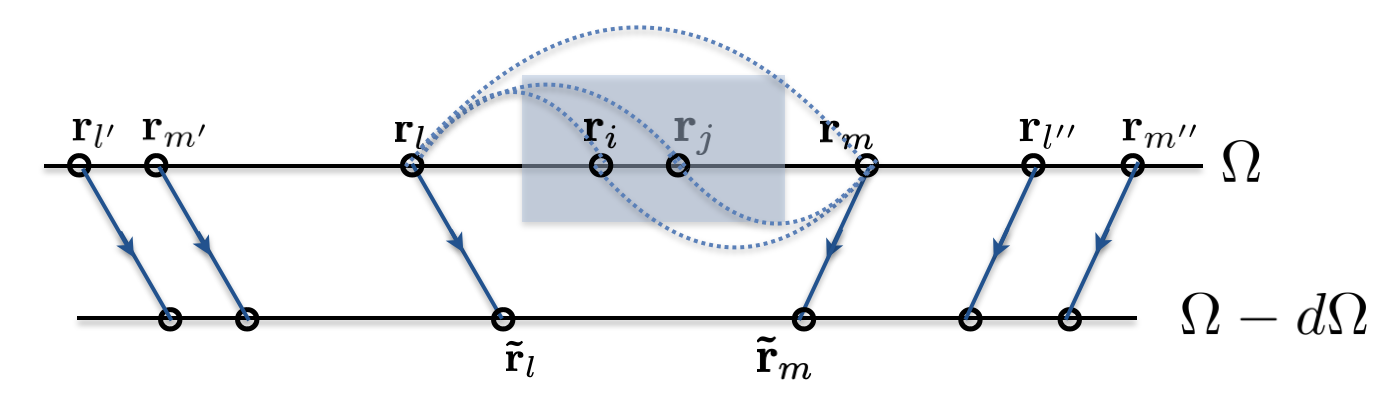}
\vspace*{-.5cm}\caption{Strong disorder RG step
for bond disordered long range coupled spin chains: Decimation of strongest coupled spin pair $(i,j)$,
highlighted by the shaded area, 
whose coupling defines the  RG scale $\Omega.$ It
is followed by  renormalization of the positions of spins,  ${\bf r}_{l} \rightarrow \tilde{{\bf r}}_{l}$ and a reduction of the RG scale to $\Omega - d\Omega.$ 
 The initial couplings   are 
indicated by the blue dashed lines. 
} 
    \label{RG}
\end{figure}

Next, we 
  apply  the  SDRG-X
             procedure 
             to the  spin chain 
           with power law long range XX-couplings,  
          Eq.(\ref{H}).

           \subsection{SDRG-X Rules}
          
           The strongest coupling  
              $J_{ij} = \Omega,$
              highlighted
               in the  example of randomly placed spins
 in Fig.  \ref{RG}, 
 forms
            one of the four pair states $|s_{ij} \rangle$.
             The 
             couplings between all remaining pairs of  spins $(l,m)$ is renormalized
according to  Eq.  (\ref{heff}),           
             depending on that state $|s_{ij} \rangle$.  
             When 
             the pair is in the singlet state $|0_{ij} \rangle$ the
             renormalized coupling is given by\cite{Mohdeb2022} 
              \begin{eqnarray} \label{jeff0}
              \tilde{J}^{}_{lm}[0_{ij}] &=&   J_{lm} - \frac{(J_{li}-J_{lj})(J_{im}-J_{jm})}{J_{ij}}.
              \end{eqnarray}

The 
  Hilbert space of 
 unentangled triplet states
             $|1_{ij} \rangle =|\uparrow_i\,\uparrow_i\rangle$ and  $|2_{ij} \rangle  =|\downarrow_i\,\downarrow_j\rangle$,
             which we denoted in Eq. 
            (\ref{heff}) by $D_1$, 
             is  
              degenerate with Eigen energy $E_1=E_2=0.$
          Deriving all terms to  second order in $V$ 
          given in 
          Eq. 
            (\ref{heff}), we find
         \begin{eqnarray} \label{heff1}
 \hat{H}_{\rm eff} [D_1] & \approx & 
 \sum_{s=1}^2 |s \rangle \langle s |
\sum_{l < m}  \tilde{J}^{}_{lm}[s] \left(S_{l}^{x}\,S_{m}^{x}+S_{l}^{y}\,S_{m}^{y}\right)
\nonumber \\
&+&  \hat{H}_{\rm new}[D_1],
\end{eqnarray}
  where  the first term contains all terms which 
  do not change the form of the interaction, resulting 
   in the renormalized interaction \cite{Mohdeb2022}
              \begin{eqnarray} \label{jeff1}
               \tilde{J}^{}_{lm}[s] &=&   J_{lm} - 
               \frac{J_{li} J_{jm} + J_{lj} J_{im}}{J_{ij}},
              \end{eqnarray}
for $s=1,2.$
  The remaining terms have a different form.
  These were not  taken into account in the implementation of 
   SDRG-X  for the long range coupled AFM XX-spin model presented in  Ref. \cite{Mohdeb2022},
   but were included recently  in Ref.\cite{Zhao2025}. 
 These terms   are  given by 
  \begin{eqnarray} \label{heff12}
 \hat{H}_{\rm new} [D_1] &=& \sum_{l < m} 
  [  (h_l S_l^z + h_m S_m^z)  \Sigma^z 
  \nonumber \\
  &+&
    \tilde{J}^{}_{lm} [1 2]  ( (S_{l}^{x}\,S_{m}^{x}-S_{l}^{y}\,S_{m}^{y})
  \Sigma^x   
   \nonumber \\
  &+& (S_{l}^{y}\,S_{m}^{x}+S_{l}^{x}\,S_{m}^{y} ) \Sigma^y ) ],
 \end{eqnarray}
 where  we introduced the pseudo spin $\Sigma$ acting on  the Hilbert space of  the 
  degenerate 2 levels, $D_1$
 with components 
% Requires: \usepackage{amsmath}
\begin{equation}
    \Sigma^x = 
    \begin{pmatrix}
        0 & 1 \\
        1 & 0
    \end{pmatrix}, 
    \Sigma^y = 
    \begin{pmatrix}
        0 & -i \\
        i & 0
    \end{pmatrix},
    \Sigma^z = 
    \begin{pmatrix}
        1 & 0 \\
        0 & -1
    \end{pmatrix}.
\label{eq:pauli_matrix_sigma_z}
\end{equation}
 The interactions between 
  the z-component of the 
  spins and the pseudospins of the degenerate 2-level system are given by
 \begin{equation} \label{hlm}
 h_l = -\frac{J_{li} J_{lj}}{J_{ij}},  h_m = -\frac{J_{im} J_{jm}}{J_{ij}},
 \end{equation}
The 3-point  interaction terms
between
the transverse components of 
2 spins and the pseudospin
are 
coupled by an effective interaction,
given by 
\begin{eqnarray} \label{jeff12}
               \tilde{J}^{}_{lm}[1,2] &=&    - 
               \frac{J_{li} J_{im} + J_{lj} J_{jm}}{J_{ij}}.
              \end{eqnarray}
Note that when keeping only the couplings between adjacent spins all 
  these  renormalized  interaction terms 
  vanish. Thus, only when taking into account 
   longer range interactions these  renormalization terms 
   are finite. 

  \begin{figure}
 \includegraphics[scale=0.18]{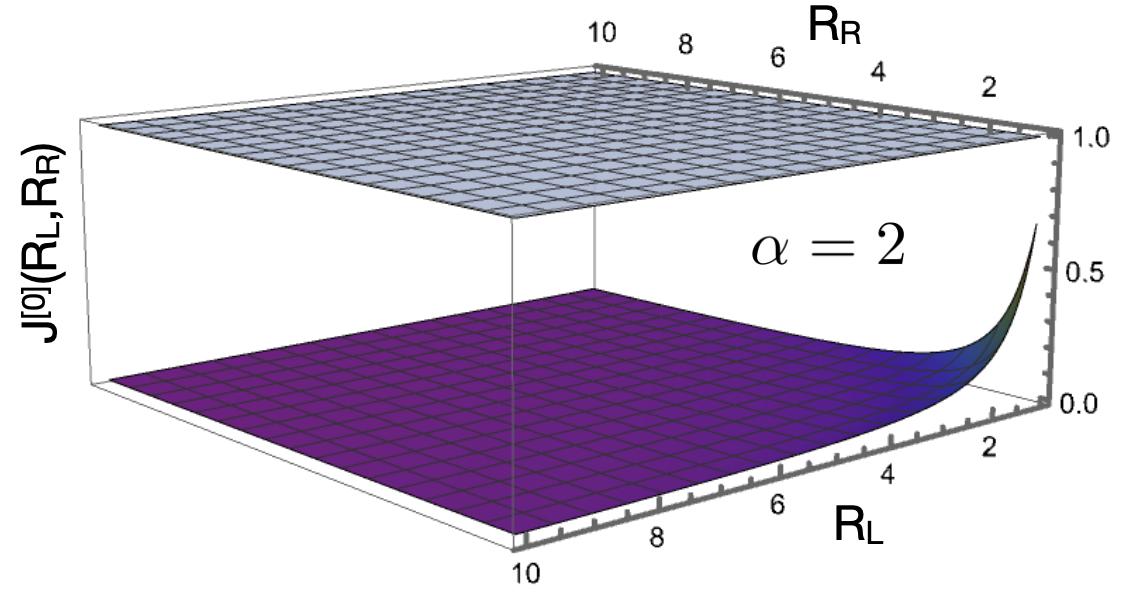}
    \includegraphics[scale=0.18]{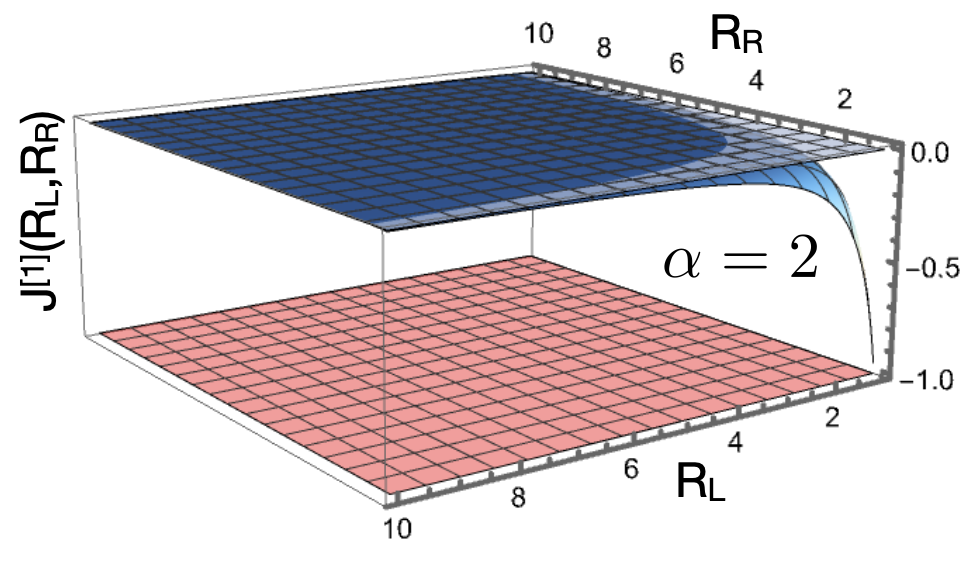}
     \includegraphics[scale=0.18]{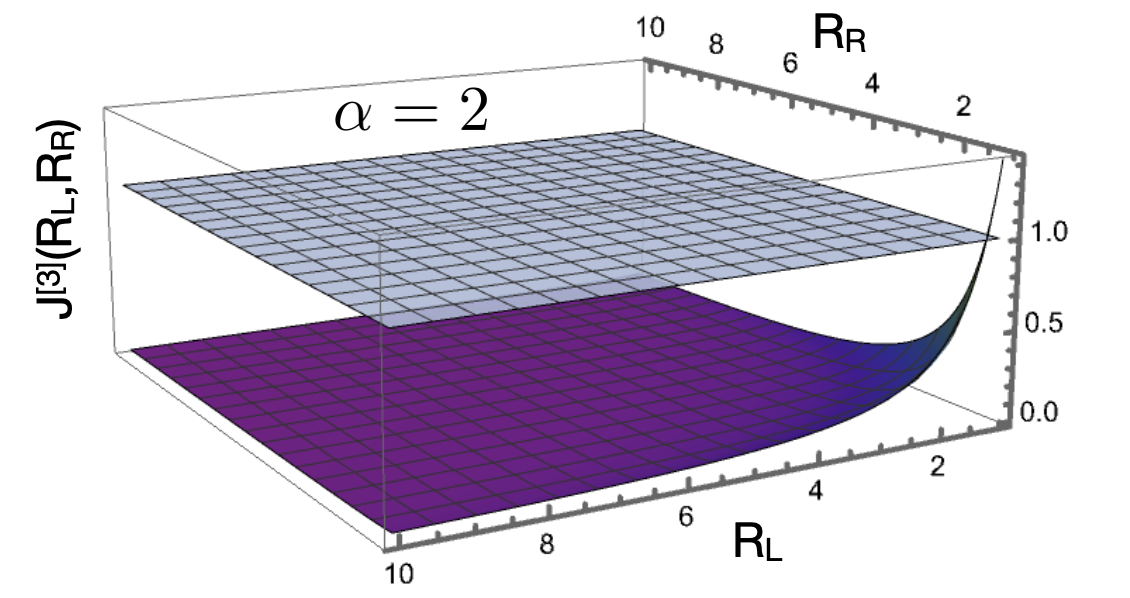}
\vspace*{-.5cm}\caption{ Renormalized couplings   $\tilde{J}[s]$ are 
plotted 
for  different projection states  $s = 0$ (upper), $s = 1,2$ (middle)
and $s = 3$ (lower)  in units of RG scale $\Omega$ as function of  distances 
$R_L=r_{li}$ and $R_R=r_{jm}$ in units of RG distance $\rho = (\Omega/\Omega_0)^{-1/\alpha}$  for $\alpha=2$.
} 
    \label{RGalpha2}
\end{figure}

\begin{figure}
  \includegraphics[scale=0.18]{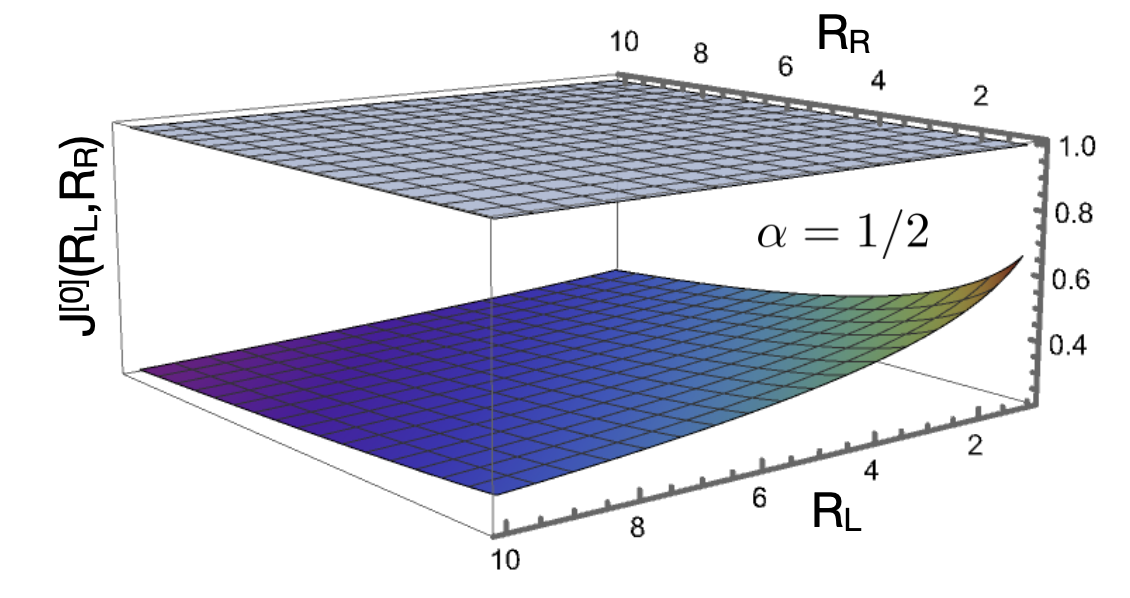}
    \includegraphics[scale=0.18]{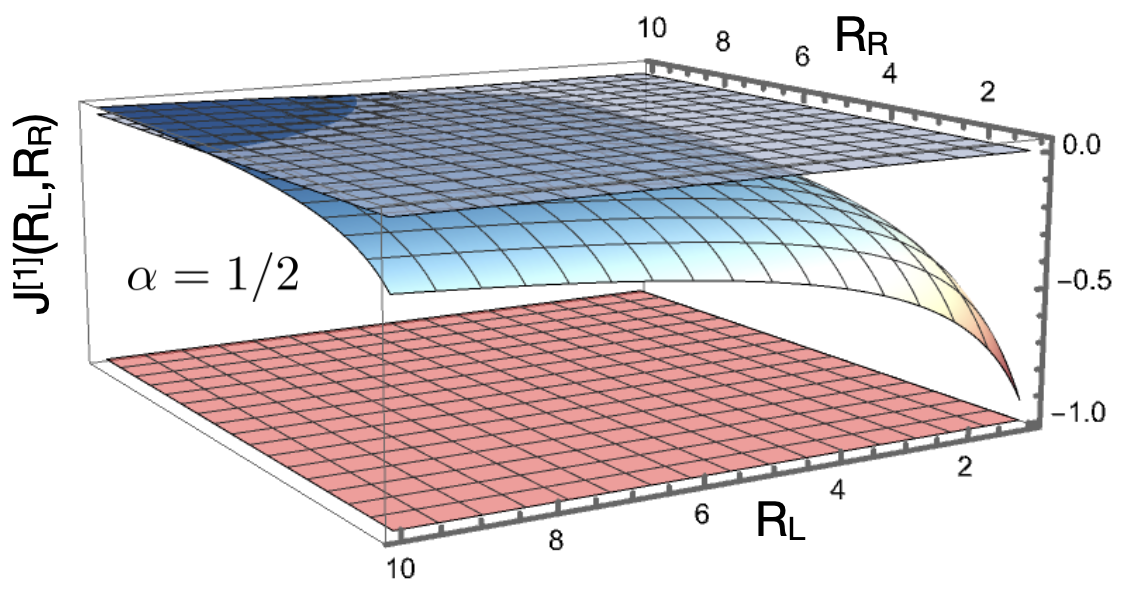}
  \includegraphics[scale=0.18]{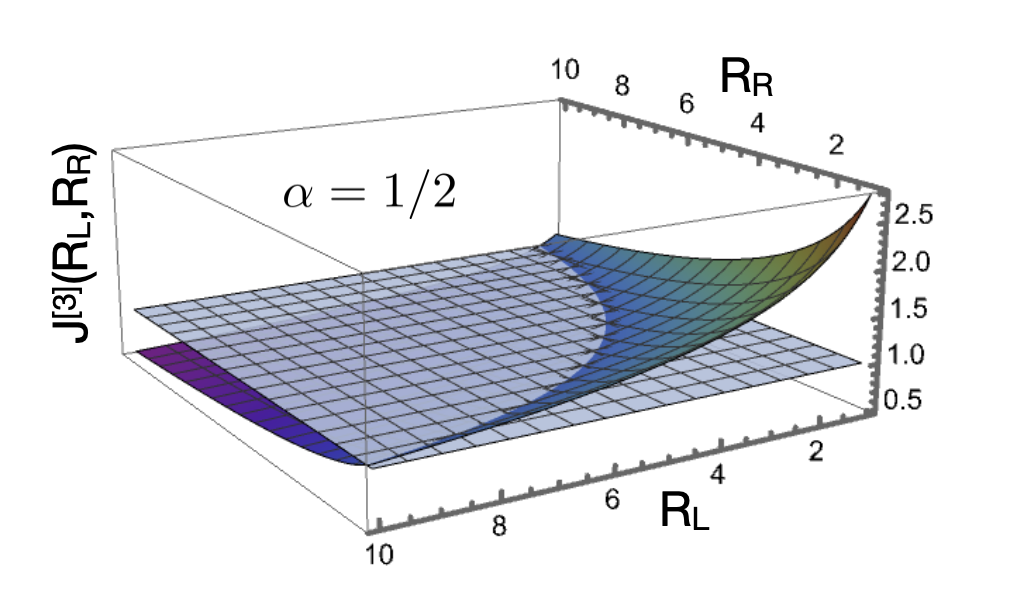}
\vspace*{-.5cm}\caption{ Same as Fig. \ref{RGalpha2} but  for $\alpha=1/2$.
} 
    \label{RGalphap5}
\end{figure}

When 
             the pair is in the entangled triplet state  $|3_{ij} \rangle$  the 
             renormalized coupling is given by\cite{Mohdeb2022}
                 \begin{eqnarray} \label{jeff3}
               \tilde{J}^{}_{lm}[3_{ij}] &=&   J_{lm} +\frac{(J_{li}+J_{lj})(J_{im}+J_{jm})}{J_{ij}}. 
              \end{eqnarray}

Let us  next explore these SDRG rules in more detail. 
              The renormalized couplings 
              between  pairs of  spins $(l,m)$ 
              depend on the initial 
              coupling between  spins
                $(l,m),$  the couplings between 
                 the removed spins $(i,j),$ and 
                the four couplings
                between spins   $(l,m),$
               and the 
                removed spins $(i,j),$
               shown schematically by the blue dashed lines in Fig.  \ref{RG}.
  These couplings
are very different 
for different projection states  $s \in \{ 0,1,2,3 \}$
and depend highly nonlinearly  on their  distances.
 We therefore plot them 
in units of RG scale $\Omega$ as function of  distances 
$R_L=r_{li}$ and $R_R=r_{jm}$ in units of distance $\rho = (\Omega/\Omega_0)^{-1/\alpha}$ between the removed spins
 in Fig. 
    \ref{RGalpha2}
for $\alpha=2$
and in Fig. 
    \ref{RGalphap5}
for  $\alpha=1/2$.

{\it Singlet State.—}
We see that, when the removed pair is in a singlet state the 
renormalized pairing $\tilde{J}^{}[0_{ij}]$,
Eq. (\ref{jeff0}) 
is smaller than the RG scale $\Omega$ 
 for all possible distances, see Figs. \ref{RGalpha2} (upper),
    \ref{RGalphap5} (upper), so that 
     this SDRG step   is well defined,
     for any $\alpha > 0$\cite{Kettemann2025}.

\begin{figure}
    \includegraphics[scale=0.25]{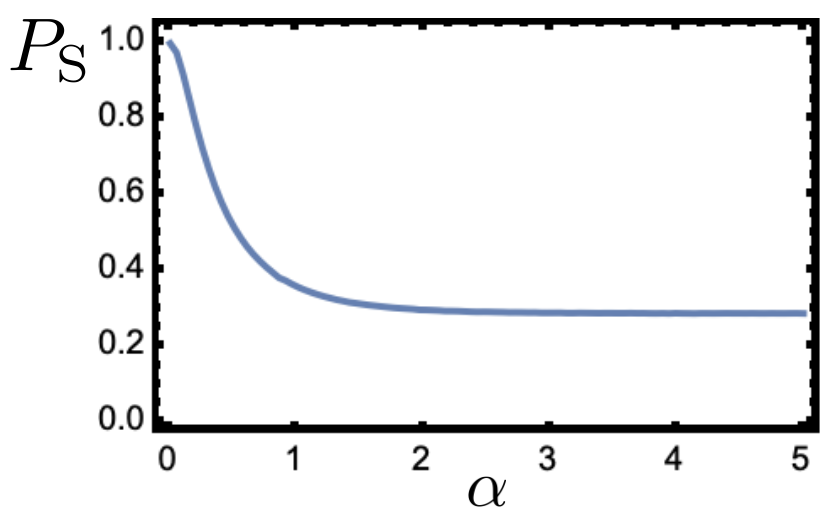}
\vspace*{-.5cm}
\caption{The probability of sign change, when 
 an unentangled triplet state forms, 
as function of $\alpha$. } 
\label{prsign}
\end{figure}

{\it Unentangled Triplet States.—}
     When the removed pair is in one of the unentangled triplet  states,  the  
renormalized pairing $\tilde{J}^{}[1_{ij}]=\tilde{J}^{}[2_{ij}]$ 
switches sign for a range of  small distances, but its amplitude remains 
smaller than the RG scale $\Omega$
 for all possible distances, see  Figs. 
  \ref{RGalpha2}(middle),
    \ref{RGalphap5}(middle). For 
    decreasing power $\alpha$ the range of distances for 
     which the sign of the  coupling changes becomes larger, 
    so that the probability of ferromagnetic renormalized  couplings increases. 
    In Fig. \ref{prsign} we plot the probability that the 
     sign changes in an RG step in which an unentangled triplet state is
      formed as function of $\alpha.$
      We see that it is finite for 
      any $\alpha$ and approaches one for $\alpha \ll 1.$  Therefore,
ferromagnetic couplings can be  encountered in 
   subsequent renormalization steps.    

 \begin{figure}
  \includegraphics[scale=0.18]{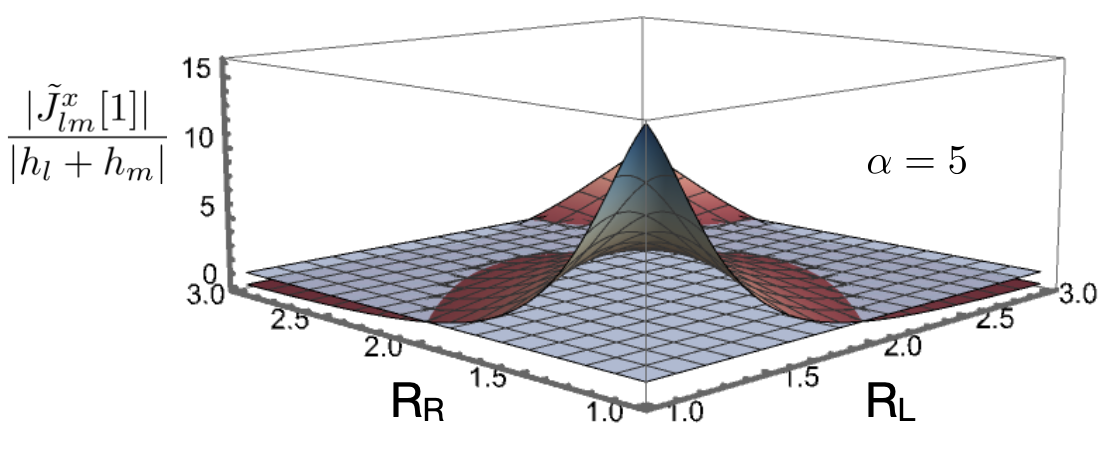}
  \includegraphics[scale=0.18]{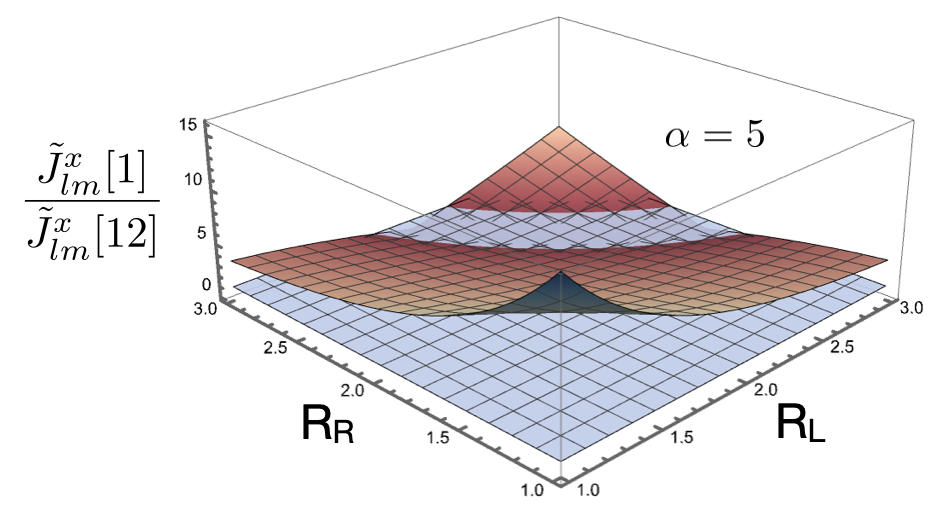}
\vspace*{-.5cm}\caption{ The ratio of the renormalized coupling 
$|\tilde{J}^{}_{lm} [1]  $
with  the sum of the  RG  generated fields  $(h_l+h_m)$
(upper figure) and 
 the new coupling strength $|\tilde{J}^{}_{lm} [1,2]$
 (lower figure), respectively, 
 as function of $R_L$ and $R_R$ in units of  $\rho$.
} 
    \label{ratio}
\end{figure}
  
    \begin{figure}
  \includegraphics[scale=0.18]{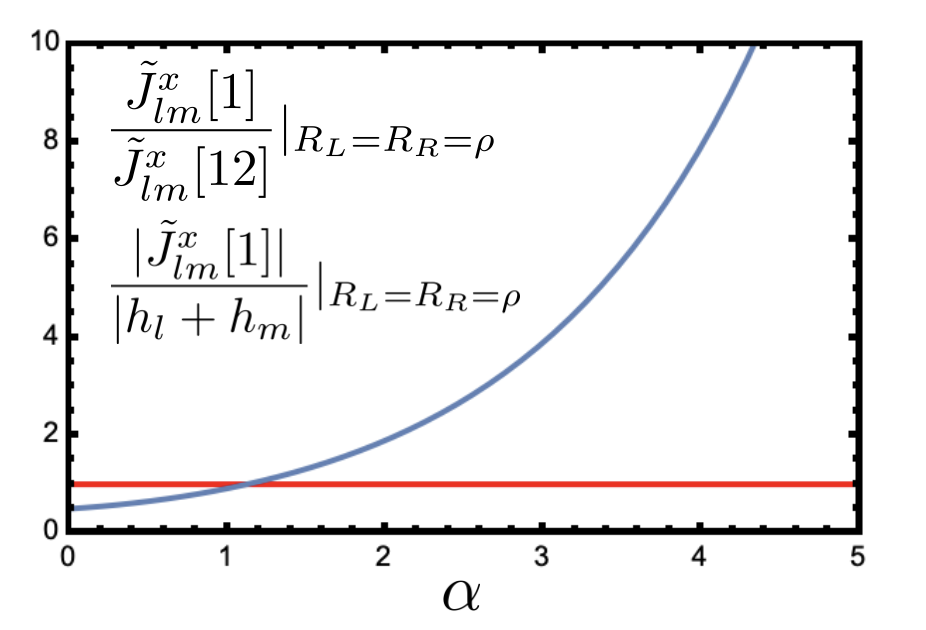}
\vspace*{-.5cm}\caption{ 
The ratio of the renormalized coupling 
$|\tilde{J}^{}_{lm} [1] | $
and  the sum of the  RG  generated  couplings $|h_l+h_m|$, and 
 the 3-point  coupling strength $|\tilde{J}^{}_{lm} [1,2]|$, respectively, 
 for $R_L=R_R= \rho.$
} 
    \label{ratior1}
\end{figure} 

However, as we reviewed above, when the 
 pair is in the degenerate space of the unentagled triplet space, also
 interactions between spins and pseudospins $  h_l,  h_m$ and 
 3-point   couplings  $ 
               \tilde{J}^{}_{lm}[1,2]$
               are generated\cite{Zhao2025}. 
            We plot the ratio of the 
             renormalized couplings which preserve
             the form of the couplings 
              $ 
               \tilde{J}^{}_{lm}[1]$
with  the sum of the local fields $  h_l + h_m$ and 
with the  new  couplings  $ 
               \tilde{J}^{}_{lm}[1,2]$
         in Fig.   \ref{ratio} (upper,lower),
         respectively.   
We see that for $R_L,R_R \rightarrow \rho,$
where the couplings are strongest, 
that ratio is maximal. 
         It turns out that at $R_L=R_R =\rho$ both ratios
         are identical and exceed
         1 for $\alpha \gg 1$, 
         so that the conventional 
         coupling dominates both the generated couplings $h_l,h_m$ and 
         $  \tilde{J}^{}_{lm}[1,2]$, as seen in Fig.
    \ref{ratior1}. 

\begin{figure}
    \includegraphics[scale=0.22]{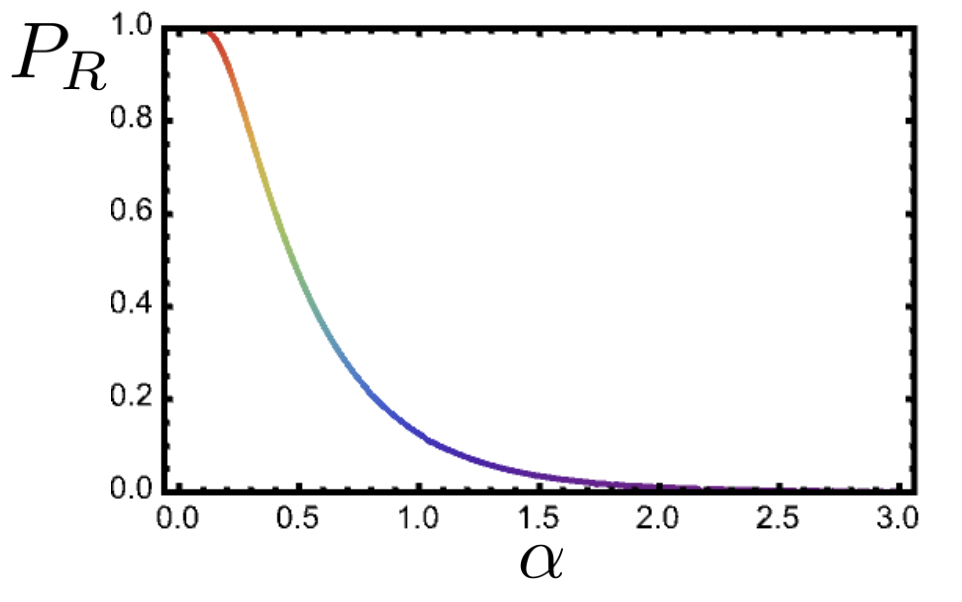}
\vspace*{-.5cm}
\caption{The probability 
that  
the renormalized distance $\tilde{r}$ becomes smaller than 
 the distance of the removed pair $\rho$
as function of $\alpha$ when the removed pair is in the entangled triplet state, $|3_{ij}\rangle$ for $\gamma=0,$ the XX-model.} 
\label{pra0}
\end{figure}

{\it Entangled Triplet State.—}
   When   the removed pair is in 
    the entangled triplet  state,  the 
renormalized pairing $\tilde{J}^{}[3_{ij}]$ remains  positive for all 
possible distances.
 However, it can 
 exceed 
the RG scale $\Omega$ for a finite  range of  distances
$R_L,R_R$
  as seen in Figs. \ref{RGalpha2},\ref{RGalphap5} (lower), 
  thus violating the consistency condition for the SDRG. 
 In Fig \ref{pra0} we plot   the probability 
that  
the renormalized distance $\tilde{r}$ is smaller than 
 the distance of the removed pair $\rho,$
as function of $\alpha,$ when the removed pair is in the entangled triplet state, $|3_{ij}\rangle$. We find that a violation 
 of the SDRG condition can occur for any $\alpha$
  and its probability increases sharply 
 for $\alpha<2.$
Indeed, in  Ref.  \cite{Mohdeb2022}, we  found 
by numerical solution of the SDRG equation and numerical exact diagonalization
that at small power exponent
$\alpha \ll 2$ the excited  states 
are no longer random pair states but rather 
become (imperfect)
rainbow states, 
where a finite number of  subsequent pairs overarch previous ones. 
This is in accordance with our 
         finding above, that,
         when the removed pair is in 
       the entangled tripled state, the renormalized  coupling 
         can be  larger  than the RG scale $\Omega$ for $\alpha <2,$ and
         accordingly, 
        the renormalized distance  smaller than the previous one
       $\tilde{r} < \rho,$  for a finite range  of distances
        of removed pairs $R_L,R_R$.
       Then, in the next RG step, 
       the following pair is forced to {\it overarch} the previous one, as in  a rainbow state.
%When the  SDRG scheme breaks down,  we need  therefore to  extend the SDRG by removing   a  larger cluster of spins and find its ground state, for example  two pairs of  spins, which form singlets, overarching each other. The renormalized couplings are then calculated using Eq. 

   Since  the sign of the couplings can switch 
    with finite probability during an RG step, 
 Fig. \ref{prsign}, 
  the sign has 
   to be traced
as function of the RG scale $\Omega.$
Therefore,  we need to define 
 the distribution function
 $P_{\sigma,T}(\tilde{r},\Omega),$
the pdf of 
the  renormalized 
distances  between adjacent spins $\tilde{r}$ at RG scale $\Omega$
 with  coupling sign $\sigma$  at temperature $T$. 
  As the renormalization proceeds to lower RG scales, 
  the  sign of the coupling  may switch frequently, as 
  all
   renormalized couplings in  Eqs. (\ref{jeff0},\ref{jeff1},\ref{jeff3})
    may  switch  their  sign. 
     When the couplings have either sign,
   all pair states can 
     result according to Eqs. (\ref{jeff0},\ref{jeff1},\ref{jeff3})
     for sufficiently small $\alpha$
     in 
     effective couplings equal or exceeding the RG scale $\Omega$
     for a finite range of distances $R_L,R_R,$
so that the SDRG condition is violated and
 the formulation of the SDRG in terms of pairs of spins
   is only valid  for  sufficiently large $\alpha \gg 1.$
    For smaller $\alpha$ the SDRG rules need to be adopted, 
     for example by considering larger clusters of spins at each RG step.  But let us first consider the regime $\alpha \gg 1,$
     where we established that the pair SDRG scheme 
     as outlined above is consistent. 

 \subsection{Finite Temperature Distribution Function}
We  derive  the Master equation 
for  the distribution function
 $P_{\sigma,T}(\tilde{r},\Omega)$
 for $\alpha \gg 1$  for $\tilde{r} \gtrapprox \rho$
 in appendix B as 
  \begin{eqnarray} \label{mlfinalc}
 - \frac{d}{d \Omega} 
 P_{\sigma,T}(\tilde{r},\Omega) &=&
P_{T}(\Omega,\Omega) P_{\sigma,T}(\tilde{r},\Omega) 
\nonumber  \\ &+&
\sum_{\sigma'}P_{\sigma',T}(\Omega,\Omega)  C_{\sigma,\sigma',T}(\tilde{r},\Omega).
      \end{eqnarray} 
Here, we used the notation
$
P_{T}(\Omega,\Omega) =
\sum_{\sigma'}P_{\sigma',T}(\Omega,\Omega). $
and we defined the function which contains 
 all terms arising due to the renormalization of the couplings when the removed pair is in 
  one of the states $s=0,1,2,3,$ with energy
  $E_{s}(\sigma \Omega)$ 
  with probability   $p_{s,T}(E_{s}(\sigma \Omega))$, 
\begin{eqnarray} \label{c}
 &&
  C_{\sigma,\sigma',T}(\tilde{r},\Omega) =
  \sum_{s=0}^3  p_{s,T}(E_{s}(\sigma' \Omega)) 
 \sum_{\sigma_L,\sigma_R,\sigma_{lm}} P_{\sigma_{lm},T}(\Omega)
 \nonumber \\
 &&
 \int_{\rho}^{\infty}
 dR_L \int_{\rho}^{\infty}
 dR_R P_{\sigma_L,T}(R_L,\Omega)  P_{\sigma_R,T}(R_R,\Omega) 
 \times 
 \nonumber \\
 &&
  ( \delta (\tilde{r}- f[s](R_L,R_R,\rho)_{ \sigma_{lm} \sigma' \sigma_L \sigma_R   } )
  ( \delta_{\sigma,\sigma_{lm}}|_{R_L,R_R \in A_{s} }
  \nonumber \\
 &&
 + 
 \delta_{\sigma, \sigma_{s} \sigma_L \sigma_R \sigma'}|_{R_L,R_R \in A^C_{s} }
) 
 \nonumber \\
 &&
 - \delta (\tilde{r}-(R_L+\rho+R_R) )\delta_{\sigma,\sigma_{lm}} ).
      \end{eqnarray}    
      Here,  in order to trace the sign of the renormalized
coupling correctly, when transforming to RG rules for the renormalized distance 
$\tilde{r}$ we
introduced the region $A_s$ of distances 
$R_L, R_R$, where the
bare coupling is larger than the renormalization correction
when the pair is in state $s,$
allowing no sign change, and its complement region $A_s^C.$
Aiming for an iterative solution, 
we first  solve the equation without the 
renormalization terms
 \begin{eqnarray} \label{mlfinalc}
 - \frac{d}{d \Omega} 
 P^0_{\sigma,T}(\tilde{r},\Omega) =
P^0_{T}(\Omega,\Omega) P^0_{\sigma,T}(\tilde{r},\Omega).
      \end{eqnarray} 
 It has the solution  
 \begin{eqnarray} \label{sdrgp0}
     P^0_{\sigma,T}(\tilde{r},\Omega) &=&  P^0_{\sigma,T}(\Omega)\frac{1}{2 \tilde{r}} (\frac{\rho}{\tilde{r}})^{1/2}  \theta(\tilde{r}-\rho),
      \end{eqnarray}
      where 
      the Bernoulli distribution of 
       the sign of the couplings is 
      $P^0_{\sigma,T} = \delta_{\sigma,+},$
      since without the renormalization all couplings are antiferromagnetic. 
Next, 
we insert  Eq. (\ref{sdrgp0}) into 
 the correction term Eq. (\ref{c}).
 Thereby the Master equation 
 Eq. (\ref{mlfinalc}) simplifies
 to 
   \begin{eqnarray} \label{mlfinalcs}
 - \frac{d}{d \Omega} 
 P_{\sigma,T}(\tilde{r},\Omega) &=&
P_{T}(\Omega,\Omega) P_{\sigma,T}(\tilde{r},\Omega) 
\nonumber  \\ &+&
Q_{\sigma,T}(\tilde{r},\Omega),
      \end{eqnarray} 
where the last term is given by 
\begin{eqnarray} \label{q}
 &&
  Q_{\sigma,T}(\tilde{r},\Omega) =
 P^0(\Omega,\Omega)  C^0_{\sigma +,T} (\tilde{r},\Omega),
      \end{eqnarray}
where the function $C^0_{\sigma +,T} (\tilde{r},\Omega)$ is defined by replacing 
 all  distribution functions in Eq. (\ref{c}) 
  by Eq. (\ref{sdrgp0}).     
The integral over one of the distances  in
 $C^0_{\sigma +,T} (\tilde{r},\Omega)$
 can now be performed, as outlined in Appendix C. 

\begin{figure}
    \includegraphics[scale=0.2]{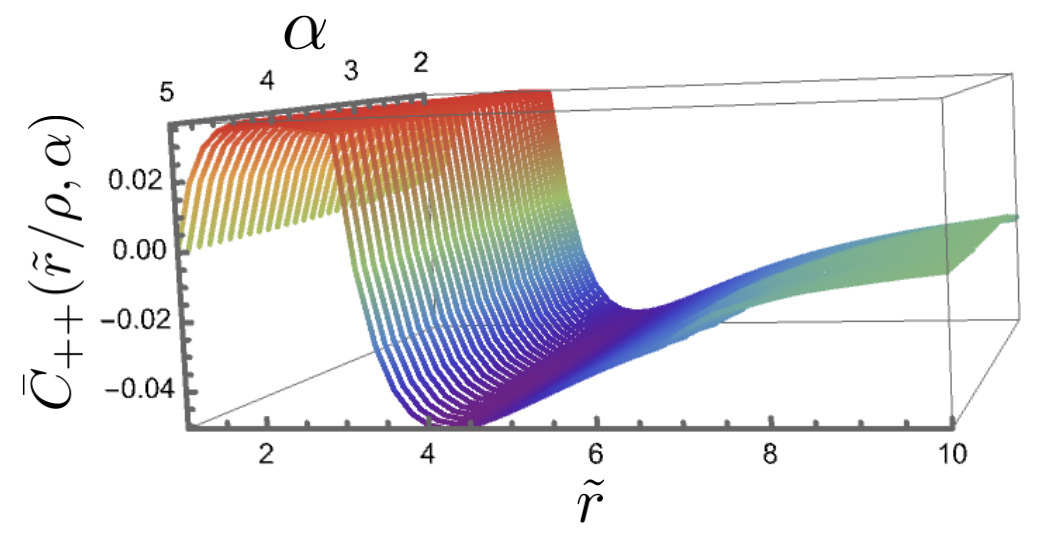}
     \includegraphics[scale=0.2]{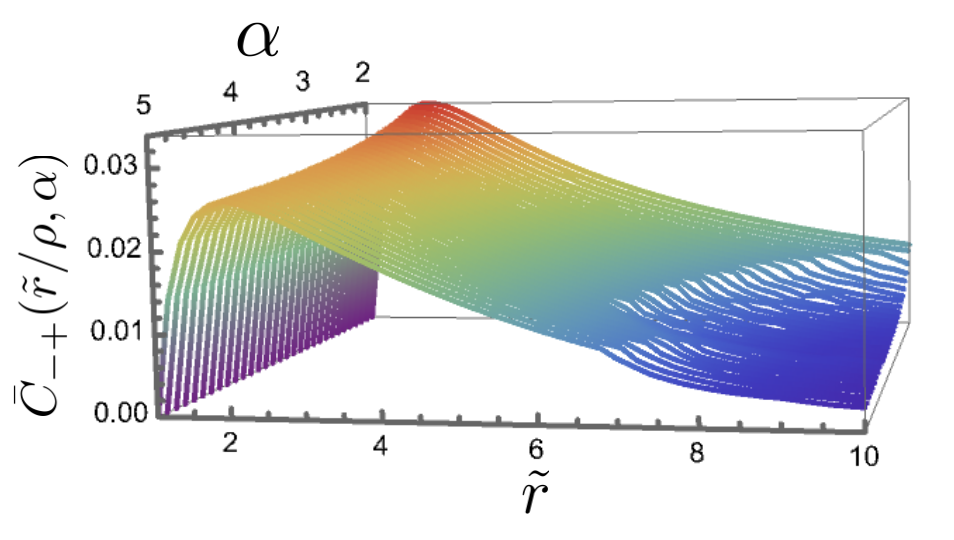}
\vspace*{-.5cm}\caption{Line plot of the correction term
in the Master equation  a) $\bar{C}_{+,+,T}(x=\tilde{r}/\rho,\alpha)$  Eq. (\ref{c3}) and b)
$\bar{C}_{-,+,T}(x=\tilde{r}/\rho,\alpha),$ 
Eq. (\ref{c4})
at finite temperature for $\Omega = 2 T$
as function of
$\tilde{r},$ for various values of power $\alpha.$  } 
    \label{corrt}
\end{figure}

We insert the  occupation probability 
     $p_{s,T}(E_{s}(\sigma' \Omega))$
     Eq. (\ref{occupation})  into 
     the resulting equation for  $C^0_{\sigma +,T} (\tilde{r},\Omega),$
     Eq.  (\ref{c2}), and 
  perform the 
integrals numerically. The result for  
   $C^0_{\sigma,+,T}(\tilde{r},\Omega \gg T)$ 
 is plotted as function of $\tilde{r}$  for various  $\alpha$
in Fig.  \ref{corrt} for $\sigma =+$(upper figure) and  $\sigma =-$(lower figure).
We see that  $C^0_{+,+,T}(\tilde{r},\Omega)$
 is  for $\alpha >2$ vanishing at $\tilde{r} =\rho$ and 
maximal for  $ \tilde{r} \approx  3\rho.$  
For  $\tilde{r}> 3\rho$  it  decays with  $\tilde{r},$ changing the sign, 
and converging to zero for  $\tilde{r} \gg 3\rho$.
$C^0_{-,+,T}(\tilde{r},\Omega)$
 is   for $\alpha >2$ also  vanishing at $\tilde{r} =\rho,$
maximal for  $ \tilde{r} \approx  2\rho,$ and decaying to zero for 
$\tilde{r} \gg 2\rho$.
 For the ground state (corresponding to $T=0K$), 
 we found in Ref. \cite{Kettemann2025} that 
  the correction  to the Master equation  is vanishing exactly at $\tilde{r}=\rho$
  for all $\alpha.$
  We find here  that also at finite temperature $T$ 
 for $\alpha \gg 2$
   the functions  $C^0_{+,+,T}(\tilde{r},\Omega)$ and   $C^0_{-,+,T}(\tilde{r},\Omega),$ 
defined in Eqs. (\ref{c3},\ref{c4}),
  are   finite  for $\tilde{r}>\rho,$ only.
 Rewriting 
   $C^0_{\sigma,+,T}(\tilde{r},\Omega) = \frac{1}{\rho}
   \bar{C}_{\sigma,+,T}(x=\tilde{r}/\rho)
   $ we solve 
  the 
  Master equation, a 1st order inhomogeneous, ordinary differential equation, as outlined in Appendix C, and 
  find that its solution is given by 
  \begin{eqnarray} \label{sdrgprp}
     P_{+}(\tilde{r},\Omega) &=&  \frac{1}{2 \tilde{r}} (\frac{\rho}{\tilde{r}})^{1/2}  \theta(\tilde{r}-\rho) \times
     \nonumber  \\ &&
     (1- \int^{\tilde{r}/\rho}_1 d x' \sqrt{x'}
     \bar{C}_{+,+,T}(x'),
      \end{eqnarray}
      and
        \begin{eqnarray} \label{sdrgprm}
        P_{-}(\tilde{r},\Omega) &=&  -\frac{1}{2 \tilde{r}} (\frac{\rho}{\tilde{r}})^{1/2}  \theta(\tilde{r}-\rho) \times
        \nonumber \\
      &&  \int^{\tilde{r}/\rho}_1 d x' \sqrt{x'}
     \bar{C}_{-,+,T}(x')   ).
      \end{eqnarray}
      The normalization condition 
      $\sum_{\sigma = \pm} \int_{\rho}^{\infty} d \tilde{r} P_{\sigma,T}(\tilde{r},\Omega) =1$
       is fullfilled exactly.
       This can 
       be seen directly  by 
         performing the integral over 
         $ \tilde{r}$ 
         with  integration  by parts
         inserting  $\bar{C}_{\sigma,+,T}(x),$ 
         using Eq. (\ref{c}), 
          performing the integral over
           the Dirac-delta functions
         and then the  sum over $\sigma.$

\begin{figure}
    \includegraphics[scale=0.2]{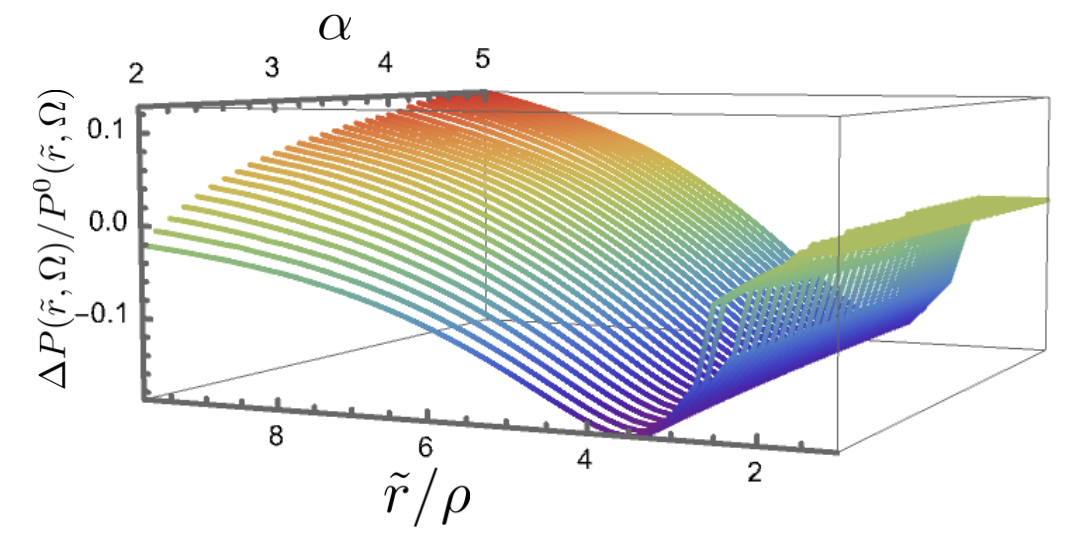}
\vspace*{-.5cm}\caption{Line plot of the ratio
 of the correction to the distribution function 
 and the SDRG distribution 
 for $\Omega = 2 T$
as function of
$\tilde{r},$ for various values of power $\alpha.$  } 
    \label{ratiodp}
\end{figure}

In Fig.  \ref{ratiodp} we plot 
the ratio
 of the correction to the distribution function 
 and the SDRG distribution $P^0 (\tilde{r},\Omega),$ 
$\delta P (\tilde{r},\Omega) =  (\sum_{\sigma} P_{\sigma}(\tilde{r},\Omega) - P^0 (\tilde{r},\Omega))/  P^0 (\tilde{r},\Omega)$ by inserting Eqs. (\ref{sdrgprp},\ref{sdrgprm})
 for $\Omega = 2 T$
as function of
$\tilde{r},$ for various values of power $\alpha.$  

    Transforming back  to the distribution of couplings
  $ P^0(\tilde{J},\Omega),$ 
  using  $ P_{\sigma,T}(\tilde{J},\Omega)|_{\tilde{J}=\Omega_0 \tilde{r}^{-\alpha} } = P_{\sigma,T}(\tilde{r},\Omega)  (\tilde{r}/(\alpha \tilde{J})),$  
  we recover the SDRG fixed point distribution Eq. (\ref{psdrg}) 
 with finite width $\Gamma = 2 \alpha$
 with small corrections to the strong disorder 
 distribution for $\tilde{J}<\Omega$. Thus, we find that   the  power law couplings  
 result in  the SDRG distribution with finite width, not only 
 for  the ground state $s=0$, as  derived in 
 Ref. \cite{Mohdeb2020}, but also for finite temperature $T$ for $\alpha \gg 2$. 
These results imply that  for $\alpha \gg 2$. 
  \begin{equation} \label{pOO}
   P_{\sigma T}(\Omega,\Omega) = \frac{1}{2 \alpha \Omega} \delta_{\sigma,+},
   \end{equation}
  which is  the  SDRG fixed point result, previously obtained for the 
   ground state\cite{Moure2018,Mohdeb2020}. Here,  we 
    derived  that result 
    to be valid for any finite temperature $T,$
    as long as  $\alpha \gg  2.$

\section{Physical Properties}

Having 
derived  the   distribution function of 
 couplings at finite temperature 
 we are now all set to 
 derive thermodynamic and  dynamic properties of bond  disordered spin chains, Eq. (\ref{H}), as summarized in the following.

 The {\it magnetic susceptibility} of
  the spin chains 
 has at finite temperature two 
 terms: 1.
 spin pairs whose coupling $J$ is smaller than 
  the temperature $T$  are broken up into two 
 free spin $S=1/2$ which thereby contribute a paramagnetic Curie law susceptibility. 
 2. Spin pairs in the unentangled triplet states, which are not broken up, 
 having a coupling $J> T,$
 act as free paramagnetic spins with $S=1.$
 Thus, the total magnetic susceptibility can be written as 
 \begin{equation}
 \chi(T) = n_{S=1/2}(T)\frac{1}{T} + n_{S=1}(T) \frac{4}{T}.
 \end{equation}
 The density of paramagnetic spins with  $S=1/2,$
 $n_{\rm S=1/2} (T)$ is 
    governed by  the differential equation 
    \begin{eqnarray} \label{nfm}
 \frac{d n_{\rm S=1/2}(\Omega)}{d \Omega} = 
2 P(J= \Omega, \Omega ) n_{\rm S=1/2} (\Omega).
      \end{eqnarray} 
      Integration yields
      \begin{equation} \label{fmhalf}
   n_{\rm S=1/2}(T) \sim \exp ( 2 \int^{T} 
   d \Omega P(\Omega, \Omega )).
   \end{equation}
  
 The density of paramagnetic spins  with $S=1,$
 $n_{\rm S=1} (T)$ is given by 
   \begin{eqnarray} \label{nf}
       n_{\rm S=1}(T) \sim
       \int_{ T } d \Omega P(\Omega,\Omega)  (p_{1,T} (\Omega) +p_{2,T} (\Omega)), 
        \end{eqnarray} 
 where
the occupation probability of states $|s \rangle,$  $s=1,2$  is 
$ p_{1,T}(\Omega)=p_{2,T}(\Omega)=
1/(2+2 \cosh (\Omega/(2T))).$

 {\it Short range Coupling.---}
 Inserting the IRFP distribution  
Eq. (\ref{psdrgf}), noting that 
$ P(\Omega, \Omega) = \sum_{\sigma} P_{\sigma,T}(\Omega,\Omega) = 1/(\ln (\Omega_0/\Omega) \Omega),$ 
 we thus find 
 \begin{equation} \label{sussr}
 \chi(T) \sim  \frac{1}{\ln (\Omega_0/T)^2} \frac{1}{T} +  \frac{c}{\ln (\Omega_0/T)} \frac{1}{T},
 \end{equation}
 where $c$ is a constant. 
Thus, we conclude that the magnetic susceptibility is dominated by paramagnetic $S=1$ pairs,  the second term in Eq. (\ref{sussr}). 

       {\it Long range Coupling.---}
      Insertion of 
      $P(\Omega,\Omega)=1/(2 \alpha \Omega),$
      and performing the integration over $\Omega$
      yields  
        \begin{eqnarray} \label{nf}
       n_{\rm S=1/2}(T) \sim (T/\Omega_0)^{1/\alpha}.
        \end{eqnarray} 
    Thereby, we find the magnetic susceptibility 
  \begin{eqnarray} \label{suslr}
\chi(T) \sim T^{\frac{1}{\alpha}-1} + \frac{c}{\alpha} T^{-1},
      \end{eqnarray} 
      where $c$ is a constant. 
      Thus, we conclude that the magnetic susceptibility is dominated by paramagnetic $S=1,$ pairs, yielding the Curie  term, the second term  in Eq. (\ref{suslr}),
      whose weight is decreasing with 
      increasing $\alpha. $

{\it Distribution of Singlet Lengths.—}

 The distribution of distances  $l$ between spins which are 
 bound into  pairs 
 $P(l),$ is determined  by  
 \begin{equation}  \label{psl}
 P(l) \sim \frac{n_{S=1/2}(\Omega)}{n_0} P(\Omega,\Omega)|_{\Omega = \Omega_0 l^{-\alpha}}
 | \frac{ d}{d l}  \Omega_0 l^{-\alpha}|.
 \end{equation} 
Inserting Eqs. (\ref{fmhalf},\ref{pOO}) we find
 \begin{equation} \label{psl2}
P(l) \sim   l^{-2},
 \end{equation}
  as previously derived  for the ground state of the  XX-Model \cite{Mohdeb2020}. Here we 
 find that it to be valid  also at finite temperature $T$ for  $ \alpha  \gg 1.$

 {\it Spin Correlation Function, Concurrence.—}

When the spins in the chain are in a pure state, such as the random pair state, the concurrence between spins 
at site $i$ and site$j$ is given by the correlation function
$C_n = |\langle \psi |\sigma_i^y \sigma_j^y|\psi \rangle |_{n= |i-j|}$.  The ensemble average is 
 given both for short range \cite{fisher94,hoyos} and long range \cite{Mohdeb2020}
 disordered antiferromagnetic quantum spin chains 
  for odd $n$ by $\langle C_n \rangle \sim P(n) \sim 1/n^2.$ The typical  correlation function 
  is dominated by deviations from the random pair state, 
  and is knwon for the short range model to be given  
  by $C_{\rm typ}(n) = \langle \exp (\ln C_n) \rangle  \sim \exp(- k \sqrt{n} )$ \cite{fisher94},
  while for the long range model 
   the typical value decays more slowly as
   $C_{\rm typ}(n)  \sim n^{-2-\alpha}$ \cite{Mohdeb2020}.
Since the distribution function 
for the pair length $P(l) \sim l^{-2}$
remains valid for the random pair state
at finite temperature, these results
for the correlation function 
remain valid
at finite temperature $T.$
 
{\it Entanglement Entropy.—}
The entanglement entropy of a subsystem 
of length $n$
with the rest of the chain is for a
specific random pair state
given by  $S_n = M_T \ln 2,$ where $M_T$ is the number of singlets $s=0$ or entangled triplet states $s=3$
 at temperature $T$,  crossing the partition of the subsystem. 
  The ensemble average of the entanglement entropy is 
 accordingly given by
  $\langle S_n \rangle = 
  \langle M_T \rangle  \ln 2.$
 The average  of $M_T$
 can be derived 
from the distribution of singlet lengths $P_s(l).$
For the ground state $T=0,$
 the leading term was found to be  given by\cite{refael},\cite{hoyos},\cite{Mohdeb2020} 
$S_n  \sim \frac{1}{2} \ln 2    \int_{l_0}^n dl ~ l P(l).$ 
At finite temperature $T$ we accordingly get 
$S_n  \sim \frac{1}{2} \ln 2    \int_{l_0}^n dl ~ l P(l) p_T (\Omega_l)$,
where $p_T (\Omega_l) = \cosh (\Omega_l/(2T))/(1+\cosh (\Omega_l/(2T)))$ 
is the probability that the pair with length $l$
is in an entangled state. Here, 
$\Omega_l = \Omega_0 l^{-\alpha}.$
This yields with  Eq. (\ref{psl2})  
a logarithmic growth of entanglement entropy with 
 subsystem length $n$ with  a correction term
 \begin{equation}
S_n (T) \sim \frac{1}{6} \ln 2  ( \ln n 
-2 \int_{z_0}^{z_n} dz \frac{1}{(1+z)^2\ln z}), 
 \end{equation}
where $z_n = \exp(\Omega_n/(2T))$
and  $z_0 = \exp(\Omega_0/(2T))$.
The first term   has  the  functional form of 
the entanglement entropy of 
critical quantum spin chains\cite{calabrese}
with effective central charge  $\bar{c} = \ln 2$\cite{refael},
 as was found previously in Ref. \cite{Mohdeb2020} for the ground state of the  XX-model with power law interactions, as 
   confirmed  there by numerical exact diagonalization.
  Here, we  found a correction term at 
  finite temperature $T$ for $\alpha \gg 1.$
For $T\rightarrow \infty,$ 
 the last term in the brackets becomes 
 $-(1/2) \ln n,$ so that the  effective central charge  halves to 
 $\bar{c}_{T \rightarrow \infty} = \frac{1}{2} \ln 2$.

{\it
Entanglement Entropy Growth After a  Quantum Quench.—
}
  Preparing the system  in  an unentangled
  state, such as a  Néel state
$|\psi_0\rangle=|\uparrow\downarrow\uparrow\downarrow\uparrow...\rangle$,
the entanglement dynamics
 can be 
monitored by the  time dependent
entanglement entropy of a subsystem with the rest of the system $S(t)$. 
When   entanglement is generated 
by   
singlet or  entangled  triplet state  across the partition,
the entanglement entropy
at time $t$ after the global  quench
is proportional to the number of
such pairs  formed 
at RG-scale $\Omega \sim 1/t$\cite{Vosk2013,Vosk2014,Igloi2012}.
 Neglecting the history of
 previously formed pairs, 
 the number of newly formed pairs at RG scale
 $\Omega,$
 $n_{\Omega}$ 
is
$   dn_{\Omega}= P(J=\Omega,\Omega) d\Omega$ \cite{refael}.
Substituting   Eq. (\ref{pOO}) we find 
$n_{\Omega} =  \frac{1}{2\alpha} \ln(\Omega)$.
Inserting  $\Omega\sim 1/t$ the entanglement entropy increases  with time as
 \begin{equation} \label{st}
    S(t) = S_p \frac{1}{2\alpha} \ln(t),
 \end{equation}
with the time-averaged  contribution of  pairs of spins $ S_p= 2\ln2- 1,$ when the initial state is a N\'eel state\cite{Vosk2013}. Then,  only singlet and  entangled triplet states are populated in the RSRG-t flow,   contributing equally.  This
  coincides  with 
  the result found in Ref. 
  \cite{Mohdeb2023}.
For the nearest neighbor XX spin chain with random bonds the growth  after a global quench is slower, $S(t) \sim\ln(\ln(t))$\cite{Vosk2013}.

\section{Conclusions}

We extended the recently introduced 
real space representation of the 
  strong disorder renormalization group 
for disordered short range and 
long range coupled quantum spin chains\cite{Kettemann2025} to 
 study finite temperature properties.
We find that 
the infinite randomness fixed point distribution for 
 short range interactions holds at finite temperature.
 We  
find small  corrections to the strong disorder  fixed point distribution
for long range interactions which 
     depend
      on power exponent $\alpha$.
For $\alpha \gg 1$ we find 
 that the value of the distribution 
 at $J=\Omega$ is independent of temperature. 
  We derive the resulting temperature dependence
   of the magnetic susceptibility, which 
    turns out to be dominated by paramagnetic 
    $S=1-$ spins resulting in a Curie law susceptibility. While the distribution 
    function of pair lengths, and the spin correlation function are found be the same as in the ground state, we find that the 
    entanglement entropy is diminished by a factor $1/2,$ when the temperature is raised to infinity. 

    For small values $ \alpha <2,$
    we find that  the corrections to the SDRG diverge and new terms, 
    interactions 
    between the z-components of the 
    spins and the 
    pseudospins,  acting on  the space of  degenerate unentangled triplet pair states, and
     three point couplings  between  
      the transverse components of pairs of  spins and the  pseudo spin emerge in the renormalized Hamiltonian. 
      Thus, for $ \alpha <2$ the SDRG method 
       has to be extended to include the renormalization flow of the new  couplings, see Ref. \cite{Zhao2025}.
       The 3-spin interactions  can possibly be treated with 
       an extension of the  real space renormalization group method to 
        multi-spin interactions, recently introduced in Ref. 
        \cite{igloi2025}.
       We found in 
        Ref. \cite{Mohdeb2022} with numerical methods  that 
   for $\alpha < 2,$ excited states are no longer random pair states, but rather (imperfect) rainbow states. Therefore,  
the SDRG has to be extended to take into account the formation of larger clusters of 
spins, forming  overarching rainbow states. 

The    real space representation    of the 
  strong disorder  renormalization group method, 
  introduced here, may    provide also an  approach to  study 
   disordered spin systems in  higher dimensions and  with 
   mixed sign couplings. 

{\it Acknowledgments.-} 
I acknowledge gratefully the hospitality of 
 the division of condensed matter at the Max-Planck institute for physics of complex systems and
  of the advanced study group on {\it strongly
  correlated extreme fluctuations}, lead by 
  Abbas Saberi. 

\section*{APPENDIX A: Derivation  of the  Finite Temperature Master Equation for Short Range Couplings}

Following the argumentation given in Ref. \cite{monthus}
 and adopting it to the distribution function of distances at temperature $T$, 
 $P_{\sigma,T}(r,\Omega),$ 
the distribution at RG scale $\Omega-d\Omega,$ lowered by  an infinitesimal amount  $d\Omega,$
is related to the distribution function at RG scale $\Omega$ by 
\begin{eqnarray} \label{prs}
 &&P_{\sigma,T}(\tilde{r},\Omega-d\Omega) =
 ( P_{\sigma,T}(\tilde{r},\Omega)+d\Omega 
 \sum_{\sigma' =\pm}
 P_{\sigma',T}(\Omega,\Omega)  \sum_{\sigma_L = \pm}  
 \nonumber \\
 && \int_{\rho}^{\infty} dR_L  
  \sum_{\sigma_R = \pm} \int_{\rho}^{\infty}
 dR_R P_{\sigma_L,T}(R_L,\Omega) 
      P_{\sigma_R,T}(R_R,\Omega) \sum_{s=0}^4 
  \nonumber \\
 &&     
       p_{s} (E_{s} ( \sigma' \Omega),T) 
 (
 \delta (\tilde{r}-  \frac{R_L R_R}{\rho}) 
 \delta_{\sigma,\sigma_L \sigma_R \sigma' \sigma_{s}}  -  \delta (\tilde{r}-R_L ) \times 
\nonumber \\
 &&  
 \delta_{\sigma,\sigma_L} -\delta (\tilde{r}-R_R) \delta_{\sigma,\sigma_R} ) ) (1-2d\Omega \sum_{\sigma'} P_{\sigma',T}(\Omega,\Omega))^{-1}.
      \end{eqnarray}  
Here, the second term 
       on the right hand side of Eq. (\ref{prs})
       accounts for 
        the addition of a  renormalized bond
        at distance $\tilde{r},$ 
        determined by the RG rule Eq.  (\ref{reff})
        with  the sign of the coupling $\sigma,$
         as given by the product of the signs of the three couplings 
         which enter the RG rule multiplied by $\sigma_{s},$
           accounting for the sign  introduced by the RG rule Eq. (\ref{RGall}). 
         The addition of that bond occurs with   probability $d\Omega \sum_{\sigma'} P_{\sigma',T}(\Omega,\Omega),$ which is 
     the probability to add a bond   of either sign $\sigma'$ 
      in  the RG step  of width $d\Omega$
      at temperature $T$. 
    The following two  terms take into account the removal of the two bonds with distance $R_L,$  $R_R,$ 
    and sign $\sigma_L,$ $\sigma_R,$
    respectively.
     These terms are integrated over all possible distances $R_L,$ $R_R$ exceeding $\rho,$ which is by definition 
      the smallest distance at RG step $\Omega.$
    In order to normalize the distribution function,  we need  to divide 
         the right side of Eq.   (\ref{prs})
        by $1-2d\Omega \sum_{\sigma}  P_{\sigma,T}(\Omega,\Omega)$, which is 
         the probability that bonds are not removed during  the RG step $d\Omega.$

      Next, performing the integrals over the last two delta functions and using the normalization  condition $\int_\rho^{\infty} d r \sum_{\sigma } P_{\sigma},T(r,\Omega) = 1,$
 performing in each term 
the sum over sign $\sigma,$
we find in the limit  $d\Omega \rightarrow 0$
the    Master equation 
for the short ranged model given in 
Eq. (\ref{mnn}).

\section*{APPENDIX B: Derivation of the Master Equation with Long Range Couplings}

 Here, we derive the Master equation for 
 the distribution function
 $P_{\sigma,T}(\tilde{r},\Omega),$
 at RG scale $\Omega$
 with  coupling sign $\sigma$  at temperature $T$. 
 We focus on the parameter  regime $\alpha \gg 1$, where we found above that the SDRG condition 
 $\tilde{r} \ge \rho$ is not violated.
  We aim to derive  $P_{\sigma,T}(\tilde{r},\Omega)$
  for $\tilde{r} \gtrapprox \rho,$
  where the remormalized coupling strength has
 the conventional form  
              $  \tilde{J}^{x}_{lm}[1],$ 
              Eq.(\ref{jeff1}),
as it   always 
   exceeds the  generated unconventional couplings
    $\tilde{J}^{x}_{lm}[1,2],$ Eq.(\ref{jeff12})
   and $h_m, h_l,$ Eq. (\ref{hlm}).

 As noted in Ref. \cite{Kettemann2025}, even though all spin pairs are interacting with each other, 
  the distribution of 
  the $N-1$ nearest neighbor distances in the chain contains 
   all information on all couplings,  since 
 the distances between all non adjacent spins 
 in the chain are  functions of those 
 $N-1$ adjacent distances.
When a 
spin pair  $(i,j)$ forms a state $| s \rangle$, $s \in \{0,1,2,3 \}$
 at RG scale $\Omega = J_{ij},$ corresponding to  a distance 
   $\rho = (\Omega_0/\Omega)^{1/\alpha},$
 the two  bonds between the two adjacent  spin pairs
 $(l,i)$ and $(j,m),$  shown in Fig. \ref{RG}, 
  with distances  
  $r_{l,i}=R_L$  and $r_{i,m}=R_R$
 are taken away.
    The   bare coupling  $J_{lm}$ is    then 
    renormalized into the  coupling $\tilde{J}_{lm}$.
     In the representation of distances 
      this corresponds to creating an adjacent bond with   renormalized distance $\tilde{r}_{lm}=\tilde{r},$  
replacing their previous distance 
${r}_{lm}={r},$ as indicated in  Fig. \ref{RG}.
For other adjacent 
    spin pairs like  $l',m'$  in Fig. \ref{RG},  where both 
    spins  $l',m'$ are located on 
    the same side
    of the singlet $(i,j),$ 
 their bare coupling is  
    renormalized into the  coupling $\tilde{J}_{l'm'}$.
     In the representation of distances 
      this corresponds to the  creation of an adjacent bond with  renormalized distance $\tilde{r}_{l'm'}=\tilde{r}$,
replacing their previous distance 
${r}_{l'm}'=\tilde{r}.$  
However, for such pairs the renormalization is small, of the order of 
$(\rho/R)^{2\alpha+2},$ where $R$ is the distance between the pair $l',m'$
and the removed pair $i,j.$  Therefore, only 
the renormalization of the distance of adjacent spins $(l,m)$ 
needs to be taken into account in the 
 derivation of  the Master equation for the distribution function  $P_{\sigma,T}(\tilde{r},\Omega).$

  Transforming the  RG rules Eqs. (\ref{jeff0},\ref{jeff1},\ref{jeff3})
   to RG  rules for the renormalized distance $\tilde{r}$
 when the pair of spins at sites $i,j$ form 
    the state $|s \rangle,$ we obtain 
the  RG rules listed in the following.
Here, we  note that the  signs of nearby couplings are not  independently distributed. We take this into account by  assuming that the sign of   close-by  couplings  between  spins at sites $l,i$ and the ones at  sites $l,j$ in Fig.  \ref{RG} are the same, $\sigma_L = \sigma_{li} =\sigma_{lj},$ and likewise $\sigma_R = \sigma_{jm} =\sigma_{im},$ while we assume that the signs of all other couplings shown in  Fig.  \ref{RG} are distributed independently.

           When 
             the pair is in the pair state $|s_{ij} \rangle$ 
             the              renormalized distance  $\tilde{r}$ is given by 
              \begin{eqnarray} \label{reff0}
           f[s] && (R_L,R_R,\rho)_{ \sigma_{lm} \sigma \sigma_L \sigma_R   } 
             \nonumber \\
             &&= 
            r |1+ \sigma_{lm} \sigma \sigma_L \sigma_R    g[s](R_L,R_R,\rho)|^{-1/\alpha},
              \end{eqnarray}
              with 
               the bare distance
              $r= (R_L + \rho +R_R).$
              Here, $\sigma,   \sigma_{lm}, \sigma_L, \sigma_R  $
              are the signs of the couplings $J_{ij} = \sigma \Omega,$
              $J_{lm},$ $J_L$
 and $J_R$, respectively.         
 We introduced              the  function $g[s](R_L,R_R,\rho)$ 
          defining    the renormalization of  the bare distance $r,$ 
          when the pair of spins taken away is in state $|s \rangle.$
         It can be rewritten as 
            \begin{equation}
            g[s](R_L,R_R,\rho)=\left(\frac{r \rho}{R_L R_R} \right)^{\alpha}  h[s](R_L,R_R,\rho),
            \end{equation}
           where only
           the 
           function   $h[s](R_L,R_R,\rho)$ depends  on the   state $s$, which the pair $(i,j)$ chooses:
           When the pair is in   the singlet state $|0_{ij} \rangle,$ it is given by 
               \begin{eqnarray} \label{g0}
            && h[0] (R_L,R_R,\rho) =  
          \nonumber \\
          &&(1-(1+\frac{\rho}{R_L})^{-\alpha}) (1-(1+\frac{\rho}{R_R})^{-\alpha}).
              \end{eqnarray}
 When 
             the pair is in one of the unentangled triplet states
             $|1_{ij} \rangle$ or  $|2_{ij} \rangle$
             then
               \begin{eqnarray} \label{g1}
             h[1] (R_L,R_R,\rho) =  -
              1-(1+\frac{\rho}{R_L})^{-\alpha}(1+\frac{\rho}{R_R})^{-\alpha},
              \end{eqnarray}
               When 
             the pair is in the entangled triplet state  $|3_{ij} \rangle$ then
                \begin{eqnarray} \label{g3}
  && h[3] (R_L,R_R,\rho) =  
          \nonumber \\
          &&(1+(1+\frac{\rho}{R_L})^{-\alpha}) (1+(1+\frac{\rho}{R_R})^{-\alpha}.
              \end{eqnarray}
             
Now,
 we are all set to  derive the Master equation 
 for the distribution function of $\tilde{r}$ 
 with coupling sign $\sigma$ at temperature $T:$ 
The distribution function of $\tilde{r}$ at 
 reduced energy scale $\Omega-d\Omega$ is given by  
\begin{eqnarray} \label{pr2}
 &&P_{\sigma,T}(\tilde{r},\Omega-d\Omega) =
 \left[  P_{\sigma,T}((\tilde{r},\Omega)  +  d\Omega \sum_{\sigma'} P_{\sigma',T}(\Omega,\Omega) 
  \right.
\nonumber \\
 && 
 \sum_{s=0}^4 p_{s,T}(E_{s}(\sigma' \Omega)) 
 \sum_{\sigma_L,\sigma_R,\sigma_{lm}}
 \int_{\rho}^{\infty} d R_L \int_{\rho}^{\infty} dR_R 
\nonumber \\
 && 
 P_{\sigma_L,T}(R_L,\Omega)  P_{\sigma_R,T}(R_R,\Omega)  
 \nonumber \\
 && 
 \left( \delta (\tilde{r}- f[s](R_L,R_R,\rho)_{ \sigma_{lm} \sigma \sigma_L \sigma_R   } ) 
 ( \delta_{\sigma,\sigma_{lm}}|_{R_L,R_R \in A_{s} }
 \right.
\nonumber \\
 && 
 + 
 \delta_{\sigma,\sigma_{s} \sigma_L \sigma_R \sigma'}|_{R_L,R_R \in A^C_{s} }
)  -\delta (\tilde{r}- (R_L+\rho+R_R)) \delta_{\sigma,\sigma_{lm}}
\nonumber \\
 &&  
 \left.  \left. 
-\delta (\tilde{r}- R_L) \delta_{\sigma,\sigma_L}
 -\delta (\tilde{r}- R_R) \delta_{\sigma,\sigma_R}
 \right)
\right]
\times
\nonumber \\
 && 
 \frac{1}{1-3 d\Omega \sum_{\sigma'} P_{\sigma',T}(\Omega,\Omega)}.
      \end{eqnarray}  
 The  delta-functions in the bracket account for the fact 
  that 
 one adjacent edge is created 
  between  sites $l$ and $m$ 
  with distance $\tilde{r},$ removing the one with distance $r,$ and removing the two adjacent bonds with distances 
   $R_L$ and $R_R.$    
Here, 
 we defined $\sigma_{s}$
              with $\sigma_{0}=1,$
              $\sigma_{1}= \sigma_{2} =-1$ accounting for the 
 minus  sign  in  $h[1](R_L,R_R,\rho),$  Eq. (\ref{g1}) and $\sigma_{3}=1.$
In order to trace the sign of 
 the renormalized coupling correctly,
   when transforming the  RG rules Eqs. (\ref{jeff0},\ref{jeff1},\ref{jeff3})
   to RG  rules for the renormalized distance $\tilde{r}$
 we  introduced  the region $A_{s}$ of distances $R_L,R_R$
  where the bare coupling is larger than the renormalization correction allowing no sign change, and its complement $A^C_{s}.$

 The last factor on the right side of Eq. (\ref{pr2}) is needed for normalization of the pdf, since 
 in total 3 edges are taken away. The proper normalization can be checked, by integrating both sides
  of Eq. (\ref{pr2}) over $\tilde{r}$ from $\rho$ to infinity, 
  Taylor expanding in $d \Omega$
 and  using the normalization 
  condition $\sum_{\sigma=\pm}\int_{\rho}^{\infty} d \tilde{r} P_{\sigma}(\tilde{r},\Omega) =1$
with  $\rho = (\Omega_0/\Omega)^{1/\alpha}.$
 Taking  the limit $d\Omega \rightarrow 0,$
we need to substract and add another  term 
 $d\Omega P(\Omega,\Omega) P(\tilde{r},\Omega)$, 
 in order to be able to cancel 
 the normalization factor in Eq. (\ref{pr2}) in the first term.
Thereby, we  find in  the limit $d\Omega \rightarrow 0,$
the Master equation  
     Eq.  (\ref{mlfinalc}) with correction term 
      given by  Eq.  (\ref{c}) .

\section*{APPENDIX C: Solution of the Master Equation with Long Range Couplings}

 In the  correction term in the Master equation Eq. (\ref{c}) we  
 perform  next the integral over distance $R_R$, 
and obtain
\begin{eqnarray} \label{c2}
 &&
  C_{\sigma,\sigma',T}(\tilde{r},\Omega) =
  \sum_{s=0}^3  p_{s,T}(E_{s}(\sigma' \Omega)) 
 \sum_{\sigma_L,\sigma_R,\sigma_{lm}}  P_{\sigma_{l,m},T}(\Omega) 
 \nonumber \\
 &&
 \int_{\rho}^{\infty}
 dR_L P_{\sigma_L,T}(R_L,\Omega)  P_{\sigma_R,T}(R[s](R_L,
 \tilde{r},\rho)_{ \sigma_{lm} \sigma' \sigma_L \sigma_R   } ,\Omega) 
  \times 
 \nonumber \\
 &&
 |\frac{d f[s]}{d R_R}|_{R_R=R[s],f[s]=\tilde{r}}|^{-1}
 \times 
 \nonumber \\
 &&
  ( \delta_{\sigma,\sigma_{lm}}|_{R_L,R[s] \in A_{s} }+ 
 \delta_{\sigma, \sigma_{s}\sigma_L \sigma_R \sigma'}|_{R_L,R[s] \in A^C_{s} }
) ) 
  \nonumber \\
 &&
 - P_{\sigma,T}(\Omega)  \theta(\tilde{r}-3\rho) \times  
  \nonumber \\
 && \sum_{\sigma_L,\sigma_R} \int_{\rho}^{\tilde{r}-2\rho}
 dR_L
  P_{\sigma_L,T}(R_L,\Omega)  P_{\sigma_R,T}( \tilde{r}-  R_L-\rho),\Omega),
      \end{eqnarray} 
where we defined $R[s](R_L,\tilde{r},\rho)_{ \sigma_{lm} \sigma' \sigma_L \sigma_R   }$  as the solution for $R_R$ of the equation 
$\tilde{r} =f[s](R_L,R_R,\rho))_{ \sigma_{lm} \sigma' \sigma_L \sigma_R   }.$

     Next, aiming for an iterative solution 
     we insert the solution without the 
      renormalization correction, as obtained when setting
      $C_{\sigma,\sigma',T}(\tilde{r},\Omega \gg T)=0$\cite{Kettemann2025},
      Eq. (\ref{sdrgp0})
      into  Eq. (\ref{c2}), yielding 
      \begin{eqnarray} \label{c3}
 &&
  C^0_{+,+,T}(\tilde{r},\Omega) =
  \sum_{s=0}^3  p_{s,T}(E_{s}(\Omega)) 
 \nonumber \\
 &&
 \int_{\rho}^{\infty}
 dR_L P_{\sigma_L,T}(R_L,\Omega)\frac{\rho}{4} 
(R[s](R_L,
 \tilde{r},\rho)_{+} ,\Omega) )^{-3/2} R_L^{-3/2}
  \times 
 \nonumber \\
 &&
 |\frac{d f[s]}{d R_R}|_{R_R=R[s],f[s]=\tilde{r}}|^{-1}
  (1|_{R_L,R[s] \in A_{s} }+ 
 \delta_{+, \sigma_{s}}|_{R_L,R[s] \in A^C_{s} }
) ) 
  \nonumber \\
 &&
 -  \theta(\tilde{r}-3\rho) 
\int_{\rho}^{\tilde{r}-2\rho}
 dR_L \frac{\rho}{4} 
  R_L^{-3/2}  (\tilde{r}-  R_L-\rho)^{-3/2},
      \end{eqnarray} 
and
  \begin{eqnarray} \label{c4}
 &&
  C^0_{-,+,T}(\tilde{r},\Omega) =
  \sum_{s=0}^3  p_{s,T}(E_{s}(\Omega)) 
 \nonumber \\
 &&
 \int_{\rho}^{\infty}
 dR_L P_{\sigma_L,T}(R_L,\Omega)\frac{\rho}{4} 
(R[s](R_L,
 \tilde{r},\rho)_{+} ,\Omega) )^{-3/2} R_L^{-3/2}
  \times 
 \nonumber \\
 &&
 |\frac{d f[s]}{d R_R}|_{R_R=R[s],f[s]=\tilde{r}}|^{-1}
  (
 \delta_{-, \sigma_{s}}|_{R_L,R[s] \in A^C_{s} }
) ),
      \end{eqnarray}

It remains  to perform the integral over $R_L$ in Eqs.  (\ref{c3},\ref{c4}).
 That can analytically be done  for the last term  in Eq.  (\ref{c3}), which 
 gives $- \theta(\tilde{r}-3\rho)  (\tilde{r}-3\rho)/(\sqrt{\tilde{r}-2\rho} (\tilde{r}-\rho)^2).$ 
We integrate  the other terms  numerically.
Inserting that result for 
  $C^0_{\sigma,\sigma',T}(\tilde{r},\Omega \gg T)$
into the Master equation, 
and  using  the Ansatz $P_{\sigma, T}(\tilde{r},\Omega \gg T) = c_{\sigma} f_{\sigma}(x={\tilde{r}/\rho})/\tilde{r},$
we find that  the Master equation  reduces to 
a 1st order inhomogeneous 
 ordinary differential equation for the  function
 $f(x=\tilde{r}/\rho).$
Solving it, and transforming back to 
 $P_{\sigma, T}(\tilde{r},\Omega)$ 
we obtain Eqs. (\ref{sdrgprp},\ref{sdrgprm}).

\end{document}